# $H_2O$- and OH-bearing minerals in the Martian regolith: Analysis of 1997 observations from HST/NICMOS

E.Z. Noe Dobrea[1], J.F. Bell III[1], M.J. Wolff[2], and K.D. Gordon[3]

[1]Cornell University, Department of Astronomy, Ithaca NY 14853-6801
[2]Space Science Institute, Boulder CO 80303
[3]Steward Observatory, University of Arizona, Tucson AZ 85721

Correspondence author contact information:

E.Z. Noe Dobrea
Cornell University
Department of Astronomy
406 Space Sciences Building
Ithaca NY 14853-6801
(607) 255-4709
ezn1@cornell.edu

Running head:  Aqueous minerals on Mars

Correspondence author contact information:

E.Z. Noe Dobrea
Cornell University
Department of Astronomy
406 Space Sciences Building
Ithaca NY 14853-6801
(607) 255-4709
[ezn1@cornell.edu](mailto:ezn1@cornell.edu)

## Abstract

We have analyzed observations of the Acidalia hemisphere of Mars taken by the Hubble Space Telescope's Near-Infrared Camera Multi-Object Spectrograph (HST/NICMOS) during July of 1997 ($L_s$ = 152°, northern Martian summer).  The data consist of images at ~60 km/pixel resolution, using both narrow- and medium-band filters specifically selected to allow us to study the hydration state of the Martian surface.   Calibration was performed by comparison to Phobos-2 ISM observations of overlapping regions, and atmospheric gas correction was performed by modeling the atmosphere for each pixel using a line-by-line radiative transfer code coupled with the MOLA altimetry data.  Our results indicate the presence of at least three spectrally different large-scale (>1000 km diameter) terrains corresponding to the dark regions of northern Acidalia, the southern hemisphere classical dark terrain, and the classical intermediate terrain adjacent to southern Acidalia.  We also identified two other spectrally unique terrains, corresponding to the northern polar ice cap, and to the southern winter polar hood.  Comparisons with mineral spectra indicate the possibility of different $H_2O$– or OH–bearing (*i.e.,* hydroxides and/or hydrates) minerals existing both in northern Acidalia and in the nearby intermediate albedo terrain. Hydrated minerals do not appear to be spectrally important components of the southern hemisphere dark terrains imaged by HST in 1997.

**Introduction**

There is considerable interest in studying the past and present stability of liquid water on the surface of Mars.  Although liquid water is not stable at the surface under today's atmospheric conditions (*e.g.,* Ingersoll, 1970; Hecht, 2002), there is significant geologic, meteoritic, and spectroscopic/isotopic evidence that liquid water may have been much more abundant on the surface and in the subsurface earlier in martian history, that it has at least sporadically flowed on the martian surface, and that it may even still be present in the subsurface today (*e.g.,* Feldman *et al.*, 2002; Boynton *et al.,* 2002; Mitrofanov *et al.*, 2002; Costard *et al.*, 2002; Malin and Edgett, 2000; Carr, 1996; Squyres *et al.*, 1992; Carr *et al.*, 1977).  However, the ambiguity of interpreting certain geomorphic features makes it difficult to determine whether or not standing bodies of water ever existed stably on the surface for geologically long periods of time.  As a result, spectroscopy combined with our terrestrial understanding of aqueous alteration, has become an important method for addressing this issue.

In contrast to the Earth, where water alteration plays a central role in the chemical weathering of the surface, the martian surface is a hyperarid environment in which liquid water cannot currently exist.  This difference alone implies that, given similar initial volcanic rocks and minerals, the subsequent mineralogical surface evolution of the two planets must differ.  Because they only require the presence of liquid water, hydrates and hydroxides are the most readily formed water-alteration products, and are very common on the Earth (*e.g.,* Salisbury *et al.*, 1991; Farmer, 1974).   Since the present atmospheric conditions on the surface of Mars preclude their formation (Gooding, 1978), their presence on the surface of the planet could be indicative of different past surface and/or subsurface environmental conditions where surface water was stable.  This hypothesis provides a compelling motivation for the search for hydration and hydroxylation products on Mars.

The near-infrared portion of the spectrum (NIR; approximately 1 to 4 µm) is the ideal region to search for $H_2O$- and OH-bearing minerals, as it contains the richest assortment of H-O-H bending and stretching bands and other O–H related spectral features (*e.g.*, Farmer, 1974; Salisbury *et al.*, 1991; Roush *et al.*, 1993; Bell, 1996). This portion of the spectrum of the water molecule is primarily characterized by a strong absorption feature at about 3-µm, caused by the overlap of the symmetric and anti-symmetric stretching modes of water with the first overtone of the strong 6-µm bending mode. Although strong, this feature does not exhibit very sensitive spectral variations with changes in mineralogy and is therefore difficult to use to constrain composition (Roush *et al.*, 1993). However, other features that involve both adsorbed and structurally bound water do exhibit diagnostic variations. These features include the 2.2 to 2.3-µm bands caused by cation-OH stretching vibrational fundamentals, the 1.9-µm H-O-H bend overtone, and the 1.4-µm OH$^-$ stretch overtone in both OH$^-$ and $H_2O$ (Salisbury *et al.*, 1991; Roush *et al.*, 1993). Because of the different physical processes responsible for these bands, it is possible to infer different properties of the host minerals from them. For example, an absorption at 1.4 µm indicates the presence of either $H_2O$ and/or OH$^-$; an absorption at 1.9 µm is an indicator of the presence of $H_2O$, such as absorbed water in smectite clays; and a 1.4-µm absorption, coupled with the absence of a 1.9-µm feature, is indicative of hydroxylated (OH-bearing) minerals, such as kaolinite.

Ground-based observations near 1.4 and 1.9 µm have been primarily plagued by the strong absorption of water in our atmosphere and, to a lesser extent, by minor absorption at 1.4-µm and major absorption at 1.9-µm due to $CO_2$ in the martian atmosphere. Due to these difficulties, ground-based observations have been restricted to the use of the 3-µm band and high-altitude observatories to determine the abundance of bound water on the surface of Mars. Sinton (1965) and, later, Beer (1971) showed evidence for a broad absorption feature around 3-μm in whole-disk observations.  Later observations of this band using the Kuiper Airborne Observatory were used to constrain the amount of bound $H_2O$ or $OH^-$ in surface materials to be about 1% to 2% (Houck *et al.,*1973).  Mariner 6 and 7 disk-resolved IRS observations were used to show the 3.1-μm to 2.2-μm band ratio to have a latitudinally-dependent distribution, with the lowest values (interpreted as increased hydration) existing in very bright areas near the equator, just north of the equator, and at latitudes below 50°S (Pimentel *et al.,*1974).  Additional IRS analyses, as well as subsequent disk-resolved ground-based (Bell and Crisp, 1993) and Phobos-2 ISM observations of the 3-μm band (Bibring *et al.,*1989; Murchie *et al.,* 1993), showed that its strength is generally correlated with albedo, although there are some dark areas exhibiting enhanced 3-μm absorptions (*e.g.,* layered deposits in the Valles Marineris; Murchie *et al.*, 2000; Calvin *et al.*, 1997).  Based on the 3-μm absorption strength, Murchie *et al.* (1993) identified two major surface units: "normal soils" with band depths of 50-60%, and "hydrated bright soils" with band depth several percent deeper than in other materials of comparable albedo.  Later, Murchie *et al.* (2000) showed that the 3-μm band is stronger in bright red soils than in most dark gray soils, but that the strongest absorptions are found in intermediate-albedo dark red soils.

The specific mineral(s) responsible for the 3-μm feature have not been identified partly because this feature does not exhibit very sensitive spectral variations with changes in mineralogy. However, spectroscopic searches for cation-OH and $OH^-$ overtone and combination features have been conducted by many researchers, often with conflicting results.  Absorption features that have been found in the region between 1.95- and 2.5-μm have been attributed to Mg-OH or amphiboles (*e.g.*, McCord *et al.*, 1978, 1982; Singer *et al.*, 1985), Fe-OH overtones (Bell and Crisp, 1993; Bell *et al.,* 1994a), and phyllosilicates (Murchie *et al.*, 1993), as well as to carbonate- or sulfate-bearing minerals (Clark *et al.*, 1990; Bell *et al.*, 1994a).

One important caveat is that many of the spectral features observed on Mars that occur at or near wavelengths characteristic of OH-bearing minerals are much narrower than their counterparts in terrestrial samples. Burns (1993b) and others have shown that one effect of dehydroxylation under desiccating conditions like those on Mars today could be to reduce the strengths and possibly, under certain circumstances, the widths of both $OH^-$ and cation-OH absorption features in the NIR. Although Bishop and Pieters' (1995) Mars soil analog experiments did not find evidence for changes in the nature of structural $OH^-$ absorption features with variation in pressure and temperature appropriate for the Martian environment, they did find evidence for systematic changes in the depth and width of $H_2O$ stretching and bending combination bands in the NIR with changes in environmental conditions and variations in the nature of interlayer smectite anionic complexes. More recently, Yet *et al.* (1999) found that UV photons bombarding Mars under current conditions are not capable of liberating $OH^-$ from mineral crystal lattices.

To take advantage of its high spatial resolution and favorable observing position above the Earth's atmosphere, we used the Near-Infrared Camera Multi-Object Spectrograph (NICMOS) on the Hubble Space Telescope (HST) to take new multispectral imaging measurements of Mars in the 1-2 µm wavelength region. In this paper we first describe the observations and provide details on their reduction and calibration. We then perform analyses of the 1.4- and 1.9-µm band depths seen in these images in order to constrain the mineralogy of the $OH^-/H_2O$-bearing host(s). This band depth analysis is compared with a similar analysis done for laboratory spectra of candidate minerals and spectral models of water-ice clouds and martian atmospheric gases, both of which could confuse or contaminate our analysis if not properly modeled. Finally, we interpret our results in terms of possible mineralogic identifications of the OH⁻/H₂O-bearing minerals on Mars, and discuss the implications for other present and future remote sensing and *in situ* compositional investigations of the planet.

**Description of the Data**

We used the HST NICMOS instrument and 11 medium- and narrow-band filters in the range between 0.95- and 2.37-μm (see Table I) to image the Acidalia hemisphere of Mars from 11:32 to 13:07 UT on July 23, 1997 (central meridian longitude ~ 35°W; $L_s$ = 152° = mid northern summer). At this time, Mars was at a phase angle of 40.1°, a geocentric distance of 1.41 AU, and a heliocentric distance of 1.53 AU. The north polar cap extended down to about 80°N latitude, and the southern polar hood extended up to approximately 40°S latitude (see Fig. 2c). The spatial resolution at the sub-earth point was 45 km/pixel, within a factor of ~2 of the Phobos2/ISM spatial resolution of ~22 km/pixel, and the highest yet achieved at these wavelengths for most of the regions in this hemisphere. The region of observation overlaps with some of the ISM tracks of classical bright and dark terrains, which allows us to directly compare datasets and hence validate our calibration. Our confidence in using the Phobos/ISM calibrated data as our standard is based on previous work where calibrated Phobos/ISM data compared well with calibrated ground-based data (Mustard and Bell, 1994; Erard and Calvin, 1997).

The NICMOS dataset consists of two sets of images taken for each filter, at 0.203 seconds total integration per set in the instrument's multiple-accumulate mode. The rationale for using the multiple-accumulate mode was that it is optimal for targets with a large dynamic range, and allows for optimal image reconstruction (deconvolution). In this mode, 4 successive images were taken during a total integration time of 0.203 seconds to generate a set, and two sets were taken for each wavelength (Fig. 1).  Despite attempts to minimize the exposure duration, many of the high albedo surface regions in the F145N filter saturated.  These regions were masked out of subsequent analysis.  Unfortunately, all the data in the F237M filter saturated, and therefore could not be used for subsequent analysis.  Use of other datasets covering the same regions (Fig. 2) aids in the calibration and analysis of our dataset.

**Data Reduction and Calibration**

<u>Instrumental Corrections</u>

The NICMOS observations were reduced using version 3.2.1 for IDL of the pipeline developed by members of the NICMOS Instrument Definition Team  (M. Reicke, private communication).  Our primary motivation for this particular version was its ability to extract useful data from parts of our saturated medium passband data (*i.e.*, "M" filters).  Direct comparisons were made between the results of the IDL-based pipeline to that of the Space Telescope Science Institute's version (*i.e.*, calnica task in IRAF/STSDAS, see http://ra.stsci.edu/STSDAS.html) with the former providing the greatest number of usable pixels.  The period between data acquisition and the reduction process allowed for the most optimal reference files to be used, with the initial photometric calibration done with conversion values kindly provided by M. Rieke (private communication, 2000).  The small number of images in our sample allowed for radiation hits to be flagged and repaired manually, with each "bad" pixel being replaced by the median of a 3x3 pixel region surrounding it.  Image reconstruction was not used for the data analysis part of this project, because of its propensity to introduce artifacts into images with large regions of oversaturated pixels (*i.e.*, the F145M filter).

Absolute photometric calibration, correction, and registration

The raw data numbers (DN) were subsequently radiometrically calibrated, co-registered, and photometrically corrected, for comparison with telescopic and Phobos/ISM data. Correction to $I/F_s$ (where I is the measured Mars irradiance and $\pi F_s$ is the input solar irradiance) was done as in Bell *et al.* (1997) using updated values for the solar flux ($F_s$) convolved to the NICMOS bandpasses (see Table I). Each image was then iteratively remapped to a Mollweide equal-area projection (as in Bell *et al.*, 1997) and manually registered until all the images were co-registered to the Viking IRTM albedo map (Palluconi and Kieffer, 1981) at a scale of 1 degree/pixel. This process also generated incidence and emission angle maps for each observation, which aided us both in the photometric and subsequent atmospheric corrections.

Photometric correction of the dataset was performed to account for limb darkening. A first-order correction was applied that assumes that the surface reflectance follows a Minnaert Law of the form

$$r_M(i, e, \Psi) = A_0 \mu_0^k \mu^{k-1}$$

where $A_0$ and k are empirical constants (the Minnaert albedo and the Minnaert parameter, respectively; *e.g.,* Minnaert, 1941; Hapke, 1993). We used a constant "typical" Minnaert parameter of k=0.7 as a good approximation to the general trend of the surface in the near-infrared (*cf.,* deGrenier and Pinet, 1995; Bell *et al.*, 1999). After photometric correction, we merged each set of image pairs into a single image for each filter. Had the photometric correction been inaccurate, we would have seen mismatches or other problems in the merging of the pairs. These image pairs also allowed us to make an estimate of the pixel-to-pixel uncertainty of the measurements and of the photometric correction for each filter.

Bootstrapping to Phobos/ISM

To test the accuracy of our reduced data, we compared our results to calibrated Phobos/ISM data of the same regions.  Our comparison shows that the calibrated NICMOS data contain systematic errors that make them inconsistent with the Phobos/ISM data (Fig. 3).  To gauge this error, we convolved the ISM data to the NICMOS bandpasses and plotted the NICMOS albedo against ISM albedo for several averaged regions for each filter (Fig. 4).  We found the error to be a systematic offset that could be represented by a combination of a wavelength-dependent multiplicative gain factor of 1.0 to 1.4 and an additive offset factor of -0.06 to -0.02 in I/F (Fig. 5).  Some of this systematic error could be related to variations in the surface and atmosphere during the ~9 year time span between the ISM and HST measurements, and some is related to the fact that the NICMOS system responsivity was not optimized or even well validated for bright object photometry.  Note, however, that the additive and multiplicative factors needed to bring the NICMOS calibration into agreement with the ISM calibration show no monotonic correlation with wavelength, which would otherwise be indicative of uncorrected photometric effects.  Note also that the gain and offset corrections are applied equally to all pixels within each image, and therefore applying these scaling factors will have no effect on multispectral ratios or other relative band-to-band derived parameters.

Correction for Atmospheric Absorption – Gaseous Component

One of the main reasons for the dearth of analyses in the 1.4- and 1.9-μm regions is the fact that water vapor absorption in our own atmosphere interferes with observations near these wavelengths.  Placing the observation platform above the atmosphere ameliorates this problem, and allows one to deal with the atmosphere of Mars alone. In order to understand and correct for the effects of Mars' atmospheric absorption in our spectra, it is necessary to model the effects of the atmosphere. These effects depend on atmospheric composition, pressure (dependent on altitude), and particle number density (which in turn depends on surface pressure and viewing geometry).

Modeling of the transmission spectrum of martian atmospheric gases was done using the line-by-line AFGL radiative transfer model of Pollack *et al.* (1993), along with the HITRAN line database developed by Rothman *et al.* (1986). In our model, we assumed a 1-layer plane-parallel atmosphere at 275 K, with 95% $CO_2$ and the nominal abundances and mixing ratios of $H_2O$ and CO given by Owen *et al.* (1992). We tested whether temperature variations of up to $\pm 50$ K have a significant effect on band depth at our spectral resolution and sampling and found that they did not (the effect of varying the atmospheric temperature by 50 K resulted in a maximum variation in transmission of 0.023 at 1.45-$\mu$m). We modeled the atmosphere for a variety of pressures (2, 4, 6, 8, 12 mbar) and particle number densities (for incidence or emission angles of 0, 45, and 75), and finally convolved these spectra to the NIMCOS bandpasses (Fig. 6).

Telltale signs of atmospheric $CO_2$ would include broad, shallow (~2%) absorption spanning the 1.45- and 1.66-$\mu$m filters, and a deeper (~5%) absorption in the 2.12 $\mu$m filter. Telltale signs of gaseous $H_2O$ would include a shallow (1-2%), broad absorption spanning the 1.13- and 1.45-$\mu$m filters, and a second absorption at the 1.90-$\mu$m filter. However, given the mixing ratios typical of Mars, it is perhaps not surprising that only the 2.12-$\mu$m $CO_2$ absorption is apparent in uncorrected NICMOS spectra (see Fig. 7).

In order to relate the modeled spectra to our observational conditions, we generated particle number density and effective pressure maps of Mars for our observations. Surface pressure maps were initially calculated using the MOLA altimetry maps (*e.g.,* Smith *et al.,* 1999), with the daytime pressure obtained by the Mars Pathfinder Lander for that time of the year (~6.7 mbar, see

Schofield *et al.,*1997) as the anchor point, and assuming an isothermal atmosphere.  Using this

method, we calculated pressures of ~1.3 mbar at the summit of Olympus Mons, ~7.2 mbar in the

interior of Valles Marineris, and 6.7 mbar at the Viking 1 landing site.  The landing site

determination agrees well with the measured pressures of this site during low dust conditions and

similar $L_s$ (VL1 P ~ 6.8 mbar; Zurek *et al.,* 1992), providing us with some confidence in the

method.  Column density $n_\alpha$ (molecules/cm$^2$) maps were created by assuming a plane-parallel

atmosphere and using

$$n_\alpha = \frac{10^3 P_s N_\alpha \alpha b}{\mu_{mol} g}$$

where $P_s$ is the surface pressure in mbar, $N_a$ is Avogadro's number, $\alpha$ is the volume mixing ratio, b is the airmass = $1/\mu + 1/\mu_0$ ($\mu_0$ = cos(incidence angle); $\mu$ = cos(emission angle)), $\mu_{mol}$ is the average

molecular mass (in grams/mol), and g is the acceleration due to gravity (also given in cgs units)

(see Appendix 1 for derivation).

Given the particle number density and the layer's pressure for each point, we bilinearly interpolated our previously derived atmospheric models to determine the atmospheric transmission spectrum for any point on the map.  However, since pressure decreases exponentially with height,

the surface pressure does not truly represent the pressure throughout the entire layer.  Therefore, we

need to define an effective pressure, $P_{eff}$, which can represent the pressure for the layer.  The

effective pressure is represented by a rescaling of the surface pressure such that

$$P_{eff} = \gamma P_{surface}$$

where $\gamma$ is a scaling factor.  To determine the value of this scaling factor, we created a set of atmospheric model maps using the values given by our particle number density map and a set of effective pressure maps whose values range from 0.05 to 1 times the surface pressure.  The depth of the 2.12-µm filter relative to the 1.66- and 2.16-µm filters ($CO_2$ absorption feature) was then used to compare the different models (using the different scaling factors) with the data, and a scaling factor of 0.45 was determined to provide the best fit (see Fig. 8).

The filters at 1.66-, 2.12-, and 2.16-µm were selected because they represent wavelengths at which the spectrum of dark terrain has a constant negative slope, as noted in the Phobos/ISM atmospherically-corrected data described by Erard *et al.* (1991).  Therefore, interpolation between the 1.66- and 2.16-µm filters allows a "continuum" value to be calculated for 2.12-µm, which can then be ratioed with the observed value to determine the observed 2.12-µm absorption depth.

Once this scaling factor was calculated, we were able to generate an atmospheric model given the effective pressure and particle number density for any point on the surface, and therefore remove (or at least characterize) the effects of atmospheric gases in the NICMOS spectrum.

The Effect of Aerosols

In order to gain some understanding on the scattering effects of aerosols on our derived band depth maps (particularly near the terminator and limb), we modeled the spectrum of aerosols in the martian atmosphere using the Discrete Ordinate Radiative Transfer (DISORT) code (Stamnes, 1988).  DISORT models multiple scattering in a layered atmosphere, and was originally developed to model scattering in the atmospheres of Earth and Mars.  The primary inputs to DISORT are the single-scattering albedo, the number of layers and optical depth of each layer, the incidence and emission angles, the azimuthal angle, and the scattering phase function of the layer. For the dust and water ice hazes of this season, our analysis focuses on the forward-scattering properties of the particles. In addition, the constant phase angle across our images minimizes the importance of the shape of the phase function for our purposes.  Consequently, we employ Mie theory to derive the single scattering albedo, which is defined as the ratio of the scattering coefficient to the extinction coefficient.  These two coefficients are in turn defined by the integral of the scattering and extinction efficiencies, respectively, weighted by the particle size distribution (n).

The scattering and extinction efficiencies are derived numerically from a modified version of the Mie scattering code of Bohren and Huffman (1983), which requires the complex refractive indices and the particle sizes as input.  We adopt the particle size distribution found in equation 2.56 in Hansen *et al.* (1974),

$$n(r) = \text{constant} \cdot r^{(1-3b)/b} e^{-r/ab}$$

where a = effective radius ($r_{eff}$) and b = effective variance ($v_{eff}$). In our case, the value of the constant is not important, because the distribution is normalized so that its integrated value equals 1.  We studied the effects of water ice and dust aerosols on our spectra.

*Water ice clouds.*  The single scattering albedo spectrum of water ice clouds was derived using the complex index of refraction published by Warren (1984) for hexagonal water-ice (*i.e.,* Ice 1h) and convolved to the NICMOS bandpasses.  In our size distribution, we assumed an effective radius of 1.5 μm with an effective variance of 0.1 (Clancy *et al.*, 2001).  The resulting single scattering albedo spectrum (Fig. 9) displays two broad and shallow (~1%) absorptions at about 1.4- and 2.0-μm (both of which are insignificant after convolving to the NICMOS bandpasses), and a deeper (~4%), sharper absorption at about 2.14-μm.  The net effect of the latter absorption is to generate a negative 1.9-μm band depth (discussed below).  This effect is observed in the southern limb of the 1.9-μm band depth map (Fig. 11), where we are looking through a greater airmass.

*Dust.* The single scattering albedo spectrum of dust was derived using a compilation of optical constants generated by Clancy *et al.* (1995). For the size distribution, we assumed an effective radius of 1.5 μm and an effective variance of 0.4 μm (Clancy *et al.,* 2001). The spectrum (Fig. 9) is almost monotonically increasing between 1.1- and 2.2-μm. Therefore, if the atmospheric dust has the optical properties of our modeled dust, it should have no discernable effects on our derived NICMOS band depth maps.

**Analysis Methods**

Band depths:

Band depth analysis is a simple yet powerful tool to explore the spatial variability of selected absorption features and to constrain the mineralogy of the surface material. Band depths are defined for each filter as

$$Band\,depth = 1 - \frac{R_0}{R_i}$$

where $R_0$ is the observed value and $R_i$ is the modeled value of the continuum at that bandpass (the continuum is modeled for each spectral band as the value derived by the interpolation of the radiance between the adjacent higher and lower wavelength bandpasses). Band depth maps were thus generated for each filter (except the two endpoints). The only case where we did not use adjacent filters to model the continuum was in the determination of the 1.9-μm band depth; for that filter the continuum value was modeled using the 1.66- and 2.14-μm filters.

Of special interest in our work are the band depths at 1.4- and 1.9-μm because of their association with hydrates and hydroxides. Scatter plots of 1.45- vs. 1.90-μm band depths of mineral spectra (obtained from the spectral library of Grove *et al.,*1992), convolved to the NICMOS bandpasses show a distribution that nominally allows us to distinguish between hydrates, hydroxides, and anhydrous minerals (see Fig. 10).  Because hydrates absorb at both 1.4- and 1.9-μm, these minerals tend to plot along a line where both band depths increase proportionally (green and purple in Fig. 10).  Hydroxides, on the other hand, absorb at 1.4-μm, and they plot along a line of varying 1.4-μm band depth and no 1.9-μm band depth (red in Fig. 10).  Finally, anhydrous minerals will tend to plot around zero band depth at both wavelengths.

Comparisons between the scatter plots in Figs. 10 and 16 and the spectroscopic and compositional information on the library minerals that we used allows us to further distinguish between groupings.  For example, most of the minerals that plot between 0.01 and 0.05 in the 1.90 μm band-depth axis and around zero in the 1.45-μm band-depth axis are primary silicates (quartz, feldspar, oxides) rather than alteration products.   Minerals exhibiting near zero 1.4 μm band-depth and near zero 1.9 μm band-depth are primarily sulfides and carbonates.  Minerals that exhibit 1.9 μm band-depths of 0.05 and greater include hydrates, sulfates, and some carbonates (*e.g.,* calcite).  Minerals plotting in the negative 1.9 μm band-depth and positive 1.4 μm band-depth region include kaolinite (aluminous) clays and micas (all containing hydroxyl), as well as a few unaltered minerals with strong near-IR iron absorption features (*e.g.,* olivine).  Most of the minerals plotting at 1.4 μm band-depths of 0.05 and greater and 1.9 μm band-depths greater than zero are plagioclases, although hydroxylated minerals plot at the highest 1.4 μm band-depths.  Minerals plotting at negative 1.4 μm band-depth (less than -0.1) and positive 1.9 μm band-depths include both hydrated and anhydrous phases, with the latter being primarily alteration products (*e.g.,* oxides like hematite) and sulfates (*e.g.,* jarosite and its variations).

<u>Principal Components Analysis:</u>

The purpose of Principal Component Analysis (PCA) is two-fold: 1) to reduce the dimensionality in our dataset without significant loss of information and 2) to identify patterns in the dataset (*e.g.,* Johnson *et al.,* 1985). In spectroscopy, PCA allows one to identify unique spectral endmembers that cannot be generated by a simple linear mixture of other spectra from the same data set. To apply PCA to our data, we first treat the data as two-dimensional matrix, where each column contains one spectrum, and each row represents a band. This matrix is then diagonalized by deriving the eigenvalues and eigenvectors of the matrix, resulting in a set of linearly independent spectra (the eigenvectors), whose contribution is defined by the eigenvalues. These endmember spectra can then be linearly mixed to attempt to model all the other spectra in the dataset.

**Error Sources and Propagation**

In our data calibration and analysis, we deal with two types of error: the first one, which we identify as the precision, is directly related to the stability of the detector and describes pixel to pixel variability induced primarily by instrumental and photometric errors. It is affected by transformations whose rescaling values vary from pixel to pixel (such as Minnaert and atmospheric corrections), but not by transformations that rescale every pixel in the array by the same amount (such as conversion from data numbers to I/F, bootstrapping to Phobos/ISM spectra, and band depth derivation). The second type of error, which we identify as the accuracy, depends on the accuracy of every value of every parameter we use in the transformation, and is therefore affected by each transformation that we apply to the data.

In our error derivation and propagation, we estimate the initial (pre-processing) precision for each filter by direct comparison of the two raw (after instrumental correction) images acquired for each filter. Using this method, we measure a standard deviation of about 1% between measurements (see Table I).

In transforming the data to I/F, our data incur some error because of inaccuracy in our knowledge of the value of the convolved solar irradiance and the scaling factor used to convert data numbers to I/F (the PHOTFLAM keyword found in the NICMOS FITS file headers). From the definition of I/F, the derived error is

$$\sigma_{I/F} = \left[ \left( \frac{-\sigma_{F_s} \times DN \times photflam \times c}{F_s^2} \right)^2 + \left( \frac{\sigma_{photflam} \times c}{F_s} \right)^2 + \left( \frac{\sigma_{DN} \times photflam \times c}{F_s} \right) \right]^{\frac{1}{2}}$$

where $\sigma$ stands for the standard deviation of the observed quantity and

$$c = \frac{\pi D^2}{4 \sin^2 \left( \Omega/2 \right)}$$

While the solar spectrum has a maximum error of 5% (Colina *et al.* 1996), we lack published uncertainties for the photflam parameter. We therefore estimate an error of 1% for this parameter, and use these values to propagate the above-derived error caused by transforming the raw data to I/F. Propagating the error through this derivation, we find our I/F to have an error in the accuracy of about 5%. Because conversion to I/F rescales the entire array equally, no error is introduced into the precision of our measurements.

The Minnaert correction in turn introduces an error of the form:

$$\sigma_{A_0} = \left[ \left( \frac{\sigma_{I/F}}{\mu_0^k \mu^{k-1}} \right)^2 + \left( \frac{I/F}{\mu^{k-1}} \frac{k \sin(i)}{\mu_0^{k+1}} \right)^2 + \left( \frac{I/F}{\mu_0^k} \frac{(k-1)\sin(e)}{\mu^k} \sigma_e \right)^2 + \left( -I/F \sigma_k \frac{\ln(\mu_0 \mu)}{\mu_0^k \mu^{k-1}} \right)^2 \right]^{\frac{1}{2}}$$

where $\mu_0$ and $\mu$ are the cosines of the incidence and emission angle, respectively, and k is the Minnaert parameter. We estimate an uncertainty of 0.5 degrees for our knowledge of the incidence and emission angles, and an uncertainty of about 0.05 (~7%) in our knowledge of the Minnaert parameter. Propagation of the errors degrades the accuracy of our data to an error of 7% and the precision to an error of about 3%.

Bootstrapping to the Phobos/ISM data introduces additional and relatively large errors in our accuracy, which we quantify from the fit we obtained when comparing NICMOS albedo with ISM albedo. The fit line has an error associated with the slope (m) and y-intercept (c), which propagates as

$$\sigma_{ism\_corr} = \left[ \left( \frac{\sigma_{A_0}}{m} \right)^2 + \left( \frac{-\sigma_c}{m} \right)^2 + \left( -\frac{A_0 - c}{m^2} \sigma_m \right)^2 \right]^{\frac{1}{2}}$$

to yield an error in the accuracy of about 21%.

Atmospheric correction also contributes to error both in the precision and accuracy. Because the atmospheric correction is a simple division, the error takes the form:

$$\sigma_{atm\_corr} = \left[ \left( \frac{\sigma_{ism\_corr}}{\mathrm{mod}el\_atm} \right)^2 + \left( \frac{-ism\_corr \times \sigma_{\mathrm{mod}el\_atm}}{\mathrm{mod}el\_atm^2} \right)^2 \right]^{\frac{1}{2}}$$

where the carbon dioxide component of the model is estimated from comparison of a large number of model spectra to be accurate to within 2% and the water vapor component has variable accuracy (because the water vapor mixing ratio varies in reality). Based on this, the atmospherically corrected data is found to conservatively have errors in the accuracy and precision of about 21% and 4%, respectively.

Because we are ultimately interested in measuring band depths, the final contributing factor to our error is the derivation of band depth maps (described above). In such a derivation, we need to consider the error that arises from interpolation as well that which arises from taking a ratio. Interpolation occurs between two filters that have errors of their own, so the error of the interpolated data is added in quadrature to the errors of the two filters being interpolated (see Bell *et al.*, 1997). The determination of band depths (a ratio) therefore causes the error in the accuracy to be about 21%. Because the transformation applies equally for all the pixels in the array, the precision is not affected and the error in the precision remains 4%.

The results of our error propagation are summarized in Table II. Not surprisingly, the greatest loss of accuracy occurs in the bootstrapping process. However, this represents the best accuracy we can obtain from our dataset if we rely on the Phobos/ISM calibration.

**Results**

Band Depths

Figure 11 presents a montage of Mollweide-projected NICMOS band depth maps, compared to a 0.97-μm albedo map of Mars. The band depth maps for the low-wavelength filters (0.97- through 1.13-μm), corresponding to the region of minimum atmospheric absorption (see Fig. 11), do not show any structure above the noise level (refer to Table III), and any possible surface variability at these wavelengths must fall below the noise. In contrast to this, the 1.45- through 1.90-μm band depth maps show strong absorptions, the strongest ones being in the 1.45- and 1.66-μm band depth maps (11%, and 5%, respectively), and corresponding to the north polar cap. In addition to the strong absorption associated with the polar cap, the 1.45-μm band depth map also shows (a) a ~3% absorption corresponding to the regions of northern Acidalia, the classical bright terrain north of Margaritifer Fossae, and the southern limb; and (b) a negative band depth (-2% to -3%) in the classical intermediate albedo regions directly southeast and southwest of Acidalia (caused by a convexity in the spectra of these regions – see Fig. 15) and in southern classical low albedo regions such as Terra Meridiani and Margaritifer Terra.  The 1.66-μm band depth map shows no other regions with band depth variability greater than 2% that correlate with specific albedo units, although variations of ~1.5% do exist between northern and southern Acidalia, and between the northern end of Margaritifer Fossae and the surrounding terrain .  Finally, the 1.9-μm band depth map shows variations of ~2% between the north polar region and the surrounding terrain, and almost 7% between the southern limb and the surrounding terrain.  There is very little or no obvious correlation to underlying geology in both cases, and all other terrain variability is of order 1% or less.

We generated 2-dimensional histogram scatter plots of the 1.45- vs. 1.9-μm band depths (Figs. 11 a and c), and the 1.45- vs. 1.66-μm band depths (Fig. 12 b and e).  The 1.9- and 1.45-μm band depths are diagnostic of hydrate and hydroxide mineralogy, whereas 1.45- and 1.66-μm band depth relations are diagnostic of the presence of water ice on the surface.   We note that for both scatter plots (with the exception of the north polar cap), most of the 1.45-μm data plots onto a main cluster ranging from –4 to 4% band depth.  The residual cap itself exhibits 1.45-μm band depths reaching ~10% (cyan region in Figs. 8 and 9).  We also note that the majority of the points not associated with the polar cap tend to cluster into 2 groups:  (Group 1) The largest group (surrounded by a yellow and a blue outline in Fig. 12) shows strong clustering around 0% band depth at 1.45-μm and 2-4% band depth at 1.90-μm as well as an anticorrelation between the 1.4- and 1.9-μm band depths.  Geographically, the data in this cluster is bound at high 1.9-μm and low 1.4-μm band depth by the intermediate albedo terrain found southeast and southwest of Acidalia Planitia, and on the low 1.9-μm and high 1.4-μm band depth by atmospheric effects near the southern limb.  The terrain encompassed by this cluster includes the classical bright regions Arabia

Terra and Tharsis, the southern (highland) classical dark regions Margaritifer Terra and Sinus

Meridiani, and  the southern portion of the low albedo region Acidalia Planitia.   (Group 2) The second cluster is centered at a relatively high 1.45-μm band depth of 2% and a 1.90-μm band depth of 3%, and does not follow the same trend as the previously described cluster.  Geographically, it encompasses the northern portion of Acidalia Planitia.

In addition to showing these two main groupings, the 1.4- vs. 1.6-μm band-depth scatter plots (Fig. 12 b and e) also indicate that the region of minimum 1.45-μm band depth is a separate group (green), independent of the previously identified Group 1, and that the region of monotonically increasing 1.45- and 1.66-μm band depth (surrounded by a blue frame) is solely associated with the residual north polar cap.

Principal Components Analysis

In order to identify spectrally unique endmembers, we derived the principal components by

finding the eigenvalues and eigenvectors in our dataset using the method described in Johnson *et al.*

(1985).  The determined eigenvalues represent the weight of each eigenvector on the data, and the

eigenvectors (Fig. 14) associated with the few highest eigenvalues contain spectral information

above the noise level.  In addition to this transformation, we generated eigenimages associated with

each eigenvector (Fig. 13), which show the geographic locations from which the eigenvector

originates.  Of particular interest are the correlations found between the band depth images and the

eigenimages, including the identification of the north polar cap and northern classical dark terrain

(*i.e.,* northern Acidalia) in the 2[nd] eigenimage, and the intermediate terrain southeast and southwest of Acidalia in the 5[th] eigenimage.

Associating the eigenvalues to the eigenimages, we find that 96.4% (the first eigenvector) of the spectral variability in our data is accounted for by albedo variations (see Figs. 8 and 9) and 3.45% (the second eigenvector) is caused mainly by the spectrum of the polar cap and, to a lesser extent, other regions. In fact, the spectrum of the second eigenvector is similar to the spectrum of water ice at most wavelengths (see Fig. 14b), and deviations from this are likely indicative of material that is mixed in with the ice and present in other bright areas of the first eigenimage. Higher eigenvectors explaining smaller percentages of the variance are correlated to different parameters (*i.e.:* band depths at 1.45, 1.66, and 1.90 μm) (see Table IV) and originate in different geographic locations (see Table V). For example, the third eigenvector shows significant correlation to the 1.9-μm band depth map (with the polar cap masked out). Data from the 6[th] eigenvector and higher are dominated by noise and do not contain any useful information.

We also created scatter plots of the eigenvectors (Fig. 15a). These types of plots produce data clouds whose extrema (or lobes) can be associated with spectral endmembers (*i.e.*, spectrally unique features that can be mixed to attempt to reproduce all the spectra in the dataset). Associating these spectral endmembers with surface features (Fig. 15b) allows us to discern real surface/atmospheric spectral endmembers from data reduction or processing artifacts. Once the endmembers are identified, we can extract and analyze their spectra from the dataset. In our analysis, we found at least 4 endmembers associated with the surface, as well as other endmembers caused by atmospheric scattering effects on these surfaces. The four surface spectral endmembers were identified by their geographic distribution as: (1) northern residual polar cap, characterized by an overall negative slope and concavity from 1.4- through 1.9-μm; (2) classical intermediate albedo terrain, characterized by a positive spectral slope up to 1.4-μm, convexity at 1.4- and 1.6-μm, concavity at 1.9-μm, and a positive slope longward of 1.9-μm; (3) southern classical dark terrains, characterized by a positive spectral slope up to 1.13-μm, spectral flatness between 1.13- and 1.66-μm, convexity at 1.66-μm, concavity at 1.90-μm, and spectral flatness longward; (4) northern Acidalia terrain, characterized by a general negative slope, with absorptions at 1.4- and 1.9-μm and spectral flatness from 1.9-μm onward.  In addition to these, we also identified an atmospheric spectral endmember associated to the southern polar hood (see Fig. 15), which is spectrally similar to that of northern Acidalia, but is convex at 1.9-μm.

**Interpretation and Discussion**

<u>Band depths</u>

As discussed earlier, band depths at 1.4- and 1.9-μm can be diagnostic of the existence of OH- and/or $H_2O$-bearing minerals.  Figures 11 and 12 show a scatter-plot comparison of our observations (blue) to measured laboratory spectra of a variety of hydrated, hydroxylated, and anhydrous minerals (described in the Analysis Methods section).  These scatter plots indicate that at the spatial resolution of our observations, there is (not surprisingly) no evidence for large (*i.e.*, over 150 km diameter) deposits of pure hydrated mineral phases on the region of Mars imaged by HST in 1997 (the cross in Fig. 16 is the precision of the data, or how well we know the location of each point with respect to each other).  Because of the accuracy of our processed data, we find that if we assume a linear mixing model and use the previously derived accuracy for our band depths, we could identify hydrates at mixing ratios of 55% or greater, but we cannot uniquely constrain the mineralogy of hydroxides at any given concentration.   Nonetheless, the detached clusters of data in Fig. 16 (more apparent in Figs. 8 and 9), which correspond to northern Acidalia and the intermediate albedo terrains, represent statistically significant and spatially contiguous deviations in 1.4 and 1.9 μm band depth.

By comparison with the pure mineral phases shown in the scatter plots (Figs. 10 and 16), we find that the main cluster of NICMOS data (corresponding to the southern dark terrains) plots in a region consistent with the presence of mostly unaltered minerals or a mixture of unaltered minerals with sulfides and/or some aluminous clays.  The band-depth cluster corresponding to higher 1.4 μm band depth and lower 1.9μm band depth (corresponding to northern Acidalia) does not plot in a region consistent with any pure hydrated mineral phase; however, it does plot along a line that could represent mixing between certain anhydrous materials (*e.g.,* felspars, oxides, plagioclases) and a subset of hydrated materials, including  sheet-silicates (chlorite), clay silicates (kaolinite, dickite), and some hydroxides (*e.g.,* brucite: $Mg(OH)_2$).  The band depth cluster corresponding to lower 1.4 μm band depth and higher 1.9μm band depth (intermediate albedo terrain) also does not plot in a region consistent with any pure hydrated mineral in our spectral library. The trend in the scatter plot is consistent, however, with a mixture of anhydrous phases and a different subset of hydrated materials, including  ferric oxyhydroxides (goethite) and oxyhydroxsulfates (jarosite).

Principal Components Analysis:

In addition to allowing us to identify 5 spectral endmembers in our data, PCA has allowed us to confirm the spectral uniqueness of the classical intermediate albedo regions southeast and southwest of Acidalia.  Figure 15c confirms the strong negative 1.4 μm band-depth identified in the intermediate albedo terrain.  The average spectra derived from the endmember analysis were compared to the mineral spectra convolved to the NICMOS bandpasses to constrain the mineralogy of the endmembers.  In our analysis, we identified mineral spectra that had similar 1.4- and 1.9-μm characteristics as those identified for the endmember spectra (see Fig. 17).  While all three non-icy surface endmembers have spectra that generally resemble the spectra of anhydrous minerals from the spectral library that we used, the spectra of northern Acidalia and of the intermediate albedo terrains have features that are consistent with the presence of hydrous minerals.  Only the southern dark terrain has a spectrum that cannot be uniquely explained by the presence of hydrous minerals.

Synthesis

Given the analysis techniques employed here and the near-IR spectral library that we used, it is not possible to constrain further the identity or abundance of the specific OH- and/or $H_2O$-bearing minerals that may be responsible for the significant variations in the near-IR spectral properties of the surface regions seen in the 1997 NICMOS data.  To better constrain possible hydrated mineral abundances will require additional analyses that explore the effects of grain size variations and systematic mixtures of hydrated minerals with other phases known or suspected to occur on Mars, especially materials like fine-grained ferric oxides that can exert a strong spectral mixing influence in the visible to near-IR (*e.g.,* Morris *et al.,* 1995). And to better constrain the possible identity of the hydrated mineral phase will require examining a much larger number of potential mineral endmembers, spanning a wider range of compositions and hydration states.

Nonetheless, band depth mapping in conjunction with PCA has allowed us to identify at least five major NIR spectral endmembers on this part of Mars.  These are the classical intermediate albedo terrains, the north polar cap, the northern Acidalia terrains, the southern classical dark terrains, and the southern polar hood.  Of particular interest are the northern Acidalia region, which is spectrally different from other dark regions in the hemisphere, and the intermediate albedo terrain, which is close to the region noted by Murchie *et al.* (2000) to have an "anomalously high" hydration state when compared to other terrains because of its enhanced 3-μm band in

Phobos-2 ISM data.

The classical northern Acidalia region exhibits a relatively high 1.45 μm band depth and

low 1.90 μm band depth, which is indicative of hydroxides when compared to similarly derived

band depths of laboratory-measured mineral spectra.  The presence of hydroxides in northern

Acidalia is in itself geologically plausible because this region is one of the lowest points on the planet and is the sink of many outflow channels that originate in the northern lowlands.  Any standing water resulting from the catastrophic outflows could have altered the surface mineralogy of the region and produced hydrates and/or hydroxides.  Whereas the current environment on Mars should have dehydrated any $H_2O$-bearing mineral phases, at least in the uppermost regolith,

hydroxides are more resistant to the desiccating conditions of the Martian atmosphere and UV

photolysis, and are more likely to have remained intact (*e.g.,* Yen *et al.,* 1999).

On the other hand, we do not find evidence for hydrated minerals either in the intermediate

terrains (as postulated by Murchie *et al.*, 2000), or anywhere else in the hemisphere observed by

NICMOS.  Our observations indicate a strong negative band depth at 1.45-μm in this terrain, which makes it distinct from other terrains in the hemisphere, but it is not consistent with the occurrence of hydrate or hydroxide minerals there.  This discrepancy could be the result of different sensitivities to mineral phases between the NICMOS and ISM data sets, or it could be related to an atmospheric aerosol or viewing geometry effect that is not being properly accounted for in one or the other data set.

Comparison with the recent Mars Odyssey Gamma Ray Spectrometer (GRS) and Neutron

Spectrometer (NS) results (Feldman *et al.*, 2002; Boynton *et al.,* 2002; Mitrofanov *et al.*, 2002)

shows consistency between the HST and Odyssey data sets for the observed terrains.  With the exception of northwestern Acidalia, GRS/NS results show a generally low hydrogen concentration in the regions identified by NICMOS as having small 1.45- and 1.90-µm band depths (consistent with a lack of hydrous minerals). In the case of northwestern Acidalia, which does show a stronger 1.45-µm absorption (indicative of hydroxides), Odyssey data also show increased hydrogen abundance.

**Conclusions**

Our analysis of calibrated and atmospherically-corrected NICMOS data has revealed a significant large-scale spectral variability on the Acidalia hemisphere of Mars. Band-depth analysis at 1.45-, 1.66-µm, and 1.90-µm has allowed us to identify 4 spectrally different terrain types: 1) north polar cap, 2) northern Acidalia, 3) southern Acidalia and southern hemisphere dark terrain, and 4) classical intermediate albedo terrain. Each of these terrains appears in different groupings in a 1.45- vs 1.90-µm band-depth scatter plot, with the northern Acidalia terrain having the highest 1.45-µm band-depth of the non-icy terrains, and the classical intermediate albedo terrain having the lowest 1.45-µm band-depth. Comparison of our NICMOS data to laboratory mineral spectra convolved to NICMOS bandpasses and analyzed similarly shows no evidence for large concentrations (greater than 55%) of hydrated minerals in this hemisphere at our resolution (~60 km/pixel). This result is not necessarily surprising, given the strong desiccating conditions of the current Martian environment. On the other hand, the segregation of the data into the three non-icy statistically significant and spatially contiguous surface components in our band depth analysis is consistent with a mineralogical diversity that may include hydroxides or other alteration products (clays, sulfates) in significant, though not yet well-constrained, fractions.

The spectral diversity of these four terrain types has been confirmed using PCA, and the extracted spectra provide evidence for a variety of possible compositions.  The spectra of the north polar cap terrain is concave from 1.45 through 1.90 μm, resembling very closely the spectrum of fine-grained water ice.  The endmember spectrum of northern Acidalia shows absorptions at both 1.45  and 1.90 μm, consistent (although not uniquely) with hydrous mineralogy.  On the other hand, the endmember spectrum of the southern dark terrain shows no evidence for absorption at 1.4 μm; its spectrum is more consistent with that of anhydrous silicate and/or oxide minerals.  Finally, the endmember spectrum of intermediate albedo terrain contains a negative 1.45 μm band and a positive 1.90 μm band, which makes it consistent with the spectra of certain hydrous and anhydrous minerals.  In addition to these four terrain types, PCA has also allowed us to identify the effect of polar hood water ice clouds on the spectrum of Mars.  This effect is primarily observed in a strong falloff of the 1.9 μm band, which is consistent with the model spectrum of the single scattering albedo of water ice.

The anhydrous nature of the southern dark terrain agrees well with the GRS results published so far, which indicate that it is a region of low hydrogen abundance.  The low hydrogen abundance of the classical intermediate terrain, as identified from GRS results, also indicates that this terrain is most likely composed of anhydrous minerals, different from those composing the southern hemisphere dark terrain.  Future NICMOS observations of the rest of the planet may allow us to directly compare the equatorial elevated hydrogen regions reported by GRS to near-infrared spectra, which will allow us to better interpret the GRS results for these regions. When combined with an expanded near-IR spectral library and more detailed spectral mixture modeling techniques, more quantitative estimates of hydrated mineral abundances should be possible.

**Acknowledgements**

We thank Richard Freedman (NASA/Ames) for earlier assistance with the use of the AFGL radiative transfer atmospheric model.  We also thank Phil James (Univ. Toledo) for assistance with the original NICMOS data acquisition, and Marcia Rieke and Dean Hines for their help in the reduction of the NICMOS data.  This work was supported by grants from the NASA Planetary Geology and Geophysics program (NAG5-10636), and a grant from the Space Telescope Science Institute, which is operated by the Association of Universities for Research in Astronomy, Incorporated, under NASA contract NAS5-26555.

# References

Beer, R., R.H. Norton, and J.V. Martonchik 1971. Astronomical infrared spectroscopy with a
  Connes-type interferometer: II-Mars, 2500-3500 cm$^{-1}$. *Icarus* **15**, 1-10.

Bell, J.F. III and D. Crisp 1993. Ground-based imaging spectroscopy of Mars in the near-infrared –
  Preliminary results. *Icarus* **104**, 1, 2-19.

Bell J.F. III, 1996. Iron, sulfate, carbonate, and hydrated minerals on Mars. In *Mineral
  Spectroscopy: A Tribute to Roger G. Burns* (Eds. M.D. Dyar, C. McCammon, and M.W.
  Scaefer), pp. 359-379. The Geochemical Society, Special Publication No. 5.

Bell, J.F. III, M.J. Wolff, P.B. James, R.T. Clancy, S.W. Lee, and L.J. Martin 1997. Mars Surface
  Mineralogy from Hubble Space Telescope Imaging During 1994-1995: Observations,
  Calibration, and Initial Results. *J. Geophys. Res.* **102**, 9109-9123.

Bibring J.P., M. Combes, Y. Langevin, A. Soufflot, C. Cara, P. Drossart, Th. Encrenaz, S. Erard, O.
  Forni, B. Bondet, L. Ksanformlity, E. Lellouch, Ph. Masson, V. Moroz, F. Rocard, and C.
  Sotin 1989. Results from the ISM experiment, *Nature* **341**, 591-593.

Bishop, J.L., C.M. Pieters, R.G. Burns, J.O. Edwards, R.L. Mancinelli, and H. Froeschl 1995.
  Reflectance spectroscopy of ferric sulfate-bearing montmorillonites as Mars soil analog
  materials. *Icarus* **117** 101-119.

Bohren, C.F. and D.R. Huffman 1983. *Absorption and scattering of light by small particles*, Wiley,
  New York.

Boynton, W. V., W.C. Feldman, S.W. Squyres, T.H. Prettyman, J. Brückner, L.G. Evans, R.C.
  Reedy, R. Starr, J.R. Arnold, D.M. Drake, P.A.J. Englert, A.E. Metzger, I. Mitrofanov, J.I.
  Trombka, C. d'Uston, H. Wänke, O. Gasnault, D.K. Hamara, D.M. Janes, R.L. Marcialis, S.
  Maurice, I. Mikheeva, G.J. Taylor, R. Tokar, and C. Shinohara 2002. Distribution of Hydrogen in the Near Surface of Mars: Evidence for Subsurface Ice Deposits. *Science* **297**, 81-84.

Burns, R.G. 1993.  Rates and mechanisms of chemical weatehring of ferromagnesian silicate materials on the surface of Mars. *Geochim. Cosmochim. Acta* **57**, 4555-4574.

Calvin, W.M. 1997.  Variation of the 3-μm absorption features on Mars: Observations over eastern Valles Marineris by the Mariner 6 infrared spectrometer.  *J. Geophys. Res.* **1002 E4**, 9097-9107.

Carr, M.H., L.S. Crumpler, J.A. Cutts, R. Greeley, J.E. Guest, and H. Masursky 1977 Martian impact craters and emplacement of ejecta by surface flow.  *J. Geophys. Res.* **82**, 4055-4065.

Carr, M.H., *Water on Mars*, Oxford Univ. Press, Oxford, 1996

Clancy, R. T., S.W. Lee, G.R. Gladstone, W.W. McMillan, and T. Rousch 1995.  A new model for Mars atmospheric dust based upon analysis of ultraviolet through infrared observations from Mariner 9, Viking, and PHOBOS.  *J. Geophys. Res.* **100, E3**, 5251-5263.

Clancy, R. T., M.J. Wolff, and P.R. Christensen 2000.  MGS TES Measurements of Dust and Ice Aerosol Behaviors.  AAS DPS meeting #32, #54.04.

Clancy, R. T., M.J. Wolff, and P.R. Christensen 2001.  Types, Sizes, Shapes and Distributions of Mars Ice and Dust Aerosols from the MGS TES Emission Phase Function Observations. AGU, Fall Meeting 2001, abstract #P42A-0542.

Clark, R.N., G.A. Swayze, R.B. Singer and J.B. Pollack 1990.  High-resolution reflectance spectra of Mars in the 2.3-μm region: Evidence for the mineral scapolite.  *J. Geophys. Res.* **95**, 14463-14480.

Colina, L., R.C. Bohlin, and F. Castelli 1996.  The 0.12-2.5 μm absolute flux distribution of the sun for comparison with solar analog stars.  *The Astronomical Journal* **112, 1**, 307-315.

Costard, F., F. Forget, N. Mangold, and J.P. Peulvast 2002. Formation of Recent Martian Debris Flows by Melting of Near-Surface Ground Ice at High Obliquity. *Science* **295, 5552**, 110-113.

Erard, S., J.-P. Bibring, J. Mustard, O. Forni, J. Head, S. Hurtrz, Y. Langevin, C. Pieters, J. Rosenqvist, and C. Sotin 1991. Spatial variations in compositions of the Valles Marineris and Isidis Planitia regions of Mars derived for ISM spectra. *Proc. Lunar Planet. Sci. Conf. XXI*, 437-455.

Erard, S. and W. Calvin 1997. New composite spectra of Mars, 0.4-5.7-μm. *Icarus* **130, 2**, 449-460.

Farmer, V.C. 1974. The layer silicates, In *The Infra-Red Spectra of Minerals* (V.C. Farmer, Ed.). The Macaulay Institute for Soil Research, London, UK

Feldman, W. C., W.V. Boynton, R.L. Tokar, T.H. Prettyman, O. Gasnault, S.W. Squyres, R.C. Elphic, D.J. Lawrence, S.L. Lawson, S. Maurice, G.W. McKinney, K.R. Moore, and R.C. Reedy 2002. Global Distribution of Neutrons from Mars Results from Mars Odyssey. *Science* **297, 5578**, 75-78.

de Grenier, M. and P.C. Pinet 1995. Near-opposition Martian limb darkening: Quantification and implication for visible-near-infrared bidirectional reflectance studies. *Icarus* **115**, 354-368.

Gooding, J.L. 1978. Chemical Weathering on Mars: Thermodynamic stabilities of primary minerals (and their alteration products) form mafic igneous rocks. *Icarus* **33**, 483-513.

Hansen, J.E. and L.D. Travis 1974. Light Scattering in Planetary Atmospheres. *Space Science Reviews*, **16**, 527-610.

Grove, C.I., S.J. Hook, and E.D. Paylor II 1992. *Laboratory Reflectance Spectra of 160 Minerals, 0.4 to 2.5 Micrometers.* JPL Publication 92-2.

Hapke, B. 1993. *Theory of Reflectance and Emittance Spectroscopy*, Cambridge University Press, New York.

Hecht, M. H. 2002. Metastability of Liquid Water on Mars. *Icarus* **156**, 373-386.

Houck, J.R., J.B. Pollack, C. Sagan, D. Schaack, and J.A. Decker Jr 1973. High Altitude Infrared Spectroscopic Evidence for Bound Water on Mars. *Icarus* **18**, 470.

Ingersoll, A.P. 1970. Mars: Occurrence of liquid water. *Science*, **168**, 972-973.

Johnson P. E., M.O. Smith, and J.B. Adams 1985. Quantitative analysis of planetary reflectance spectra with principal components analysis. *J. Geophys. Res. Supplement* **90**, C805-C810.

Malin, M.C., and K.S. Edgett 2000. Evidence for recent groundwater seepage and surface runoff on Mars. *Science* **288**, 2330-2335.

McCord T.B., Clark R.N. and R.L. Huguenin 1978. Mars: Near-infrared spectral reflectance and compositional implications, *J. Geophys. Res* **83**, 5433-5441.

McCord T.B., Clark R.N. and R.B. Singer 1982. Mars: Near-infrared spectral reflectance of surface regions and compositional implications. *J. Geophys. Res.* **87**, 3021-3032.

Minnaert, M. 1941. The reciprocity principle in lunar photometry. *Astrophys. J.* **93,** 403-410.

Mitrofanov, I., D. Anfimov, A. Kozyrev, M. Litvak, A. Sanin, V. Tret'yakov, A. Krylov, V. Shvetsov, W. Boynton, C. Shinohara, D. Hamara, and R.S. Saunders 2002. Maps of Subsurface hydrogen from the high energy neutron detector, Mars Odyssey. *Science* **297**, 5578, 78-81.

Morris R.V., D.C. Golden, J.F. Bell III, and H.V. Lauer, Jr. 1995. Hematite, pyroxene, and phyllosilicates on Mars: Implications from oxidized impact melt rocks from Manicougan Crater, Quebec, Canada, *J. Geophys. Res.* **100**, 5319-5329.

Murchie S., Mustard J., Bishop J. Head J., Pieters C. and Erard S. 1993.  Spatial variation in the spectral properties of bright regions on Mars. *Icarus* **105**, 454-468.

Murchie, S., L. Kirkland, S. Erard, J. Mustard, and M. Robinson 2000.  Near-Infrared spectral variations of Martian surface materials from ISM imaging spectrometer data. *Icarus* **147**, 444-471.

Mustard, J.F. and J.F. Bell 1994.  New composite reflectance spectra of Mars from 0.4 to 3.14 micron. *Geophysical Research Letters* **21, #5**, 353-356.

Owen, T. 1992.  The composition and early history of the atmosphere of Mars**.**  In *Mars* (H.H. Kieffer, B.M. Jakosky, C.W. Snyder, M.S. Matthews, Eds.), pp. 818-834.  Univ. of Arizona Press, Tucson.

Palluconi F.D. and H.H. Kieffer 1981.  Thermal inertia mapping of Mars from 60˚ S to 60˚ N. *Icarus* **45**, 415-426.

Pimentel, G.C., P.B. Forney, K.C. Herr 1974.  Evidence about hydrate and solid water in the Martian surface from the 1969 Mariner infrared spectrometer. *J. Geophys. Res.* **79**, 1623-1634.

Roush, T.L., D.L. Blaney, and R.B. Singer 1993.  The surface composition of Mars as inferred from spectroscopic observations.  In *Remote Geochemical Analysis: Elemental and Mineralogical Composition* (C.M. Pieters and P.A.J. Englert, Eds.), pp. 367-393, Cambridge University Press, Cambridge, UK.

Roush, T.L., J.B. Pollack, F.C. Witteborn, J.D. Bregman, and J.P. Simpson 1990.  Ice and minerals on Callisto:  A reassessment of the reflectance spectra. *Icarus* **86**, 355-382.

Salisbury, J.W., L.S. Walter, N. Vergo, and D.M. D'Aria 1991.  *Infrared (2.1-25 µm) Spectra of Minerals*.  The John Hopkins Press, Baltimore.

Schofield, J.T., J.R. Barnes, D. Crisp, R.M. Haberle, S. Larsen, J.A. Magalhães, J.R. Murphy, A.

    Seiff, and G. Wilson 1997.   The Mars Pathfinder Atmospheric Structure

    Investigaiton/Meteorology (ASI/MET) Experiment.  *Science*, **278**, 1752-1757.

Sinton, W.M. 1967.  On the composition of the Martian surface materials.  *Icarus* **6**, 222-228.

Smith, D.E., M.T. Zuber, S.C. Solomon, R.J. Phillips, J.W. Head, J.B. Garvin,W.B. Banerdt, D.O.

    Muhleman, G.H. Pettengill, G.A. Neumann, F.G. Lemoine, J.B. Abshire, O. Aharonson, D.C.

    Brown, S.A. Hauck, A.B. Ivanov, P.J. McGovern, H.J. Zwally, and T.C. Duxbury 1999. The

    Global Topography of Mars and Implications for Surface Evolution, Science **284**, 1495-1500.

Squyres, S.W., S.M. Clifford, R.O. Kuzmin, J.R. Zimbelman, and F.M. Costard 1992.  Ice in the

    martian regolith.  In *Mars* (H.H. Kieffer, B.M. Jakosky, C.W. Snyder, and M.S. Matthews,

    Eds.), pp. 523-554.  Univ. of Arizona Press, Tucson.

Stamnes, K., S. Tsay, W. Wiscombe, and K. Jayaweera 1988.  Numerically stable algorithm for

    discrete-ordinate-method radiative transfer in multiple scattering and emitting layered media,

    *Appl. Optics,* **27**, 2502-2509.

Warren, S.G. 1984.  Optical constants of ice from the ultraviolet to the microwave. *Appl. Opt.* **23**,

    1206.

Yen, A.S., B. Murray, G.R. Rossman, and F.J. Grunthaner 1999.  Stability of hydroxylated

    minerals on Mars:  A study on the effects of exposure to ultraviolet radiation.  *J. Geophys.*

    *Res.* **104, E11**, 27,031-27,041.

Zurek, R.W., J.R. Barnes, R.M. Haberle, J.B. Pollack, J.E. Tillman, and C.B. Leovy 1992.

    Dynamics of the atmosphere of Mars, in *Mars* (H.H. Kieffer, B.M. Jakosky, C.W. Snyder,

    and M.S. Matthews, Eds.), pp. 835-932. Univ. of Arizona Press, Tucson.

**Table I.**  HST NICMOS Filters and Exposure Times

| Camera | Filter | $\lambda_{center}$, µm | Target | Observation times (UT) | FWHM, µm | PHOTFLAM (W/cm²/µm)/(DN/sec) | $F_S$, W cm⁻² µm⁻¹ | $F_S$ point-t0-point variability |
|--------|--------|------|--------|------|------|------|------|------|
| NIC1 | F095N | 0.9538 | Mars | 11:32:48 | 0.0088 | 6.98300e-20 | 0.0772056 | 1.043 |
|      |       |        | Mars | 11:33:49 |        |             |           |       |
| NIC1 | F097N | 0.9717 | Mars | 11:34:50 | 0.0094 | 5.54500e-20 | 0.0763197 | 0.989 |
|      |       |        | Mars | 11:35:51 |        |             |           |       |
| NIC1 | F108N | 1.0817 | Mars | 11:36:52 | 0.0094 | 3.78700e-20 | 0.0622587 | 0.891 |
|      |       |        | Mars | 11:37:53 |        |             |           |       |
| NIC1 | F113N | 1.1297 | Mars | 11:38:54 | 0.0110 | 2.79200e-20 | 0.0568340 | 0.827 |
|      |       |        | Mars | 11:39:55 |        |             |           |       |
| NIC1 | F145M | 1.4513 | Mars | 11:40:56 | 0.1965 | 7.67000e-22 | 0.0321866 | 0.729 |
|      |       |        | Mars | 11:41:57 |        |             |           |       |
| NIC1 | F166N | 1.6606 | Mars | 11:42:58 | 0.0168 | 5.95700e-21 | 0.0231940 | 0.745 |
|      |       |        | Sky  | 11:44:20 |        |             |           |       |
|      |       |        | Mars | 11:45:41 |        |             |           |       |
| NIC1 | F190N | 1.8987 | Mars | 11:46:42 | 0.0174 | 4.04300e-21 | 0.0136482 | 0.967 |
|      |       |        | Sky  | 11:48:04 |        |             |           |       |
|      |       |        | Mars | 11:49:25 |        |             |           |       |
| NIC2 | F212N | 2.1211 | Mars | 13:04:37 | 0.0206 | 2.44500e-21 | 0.0088069 | 0.524 |
|      |       |        | Sky  | 13:05:59 |        |             |           |       |
|      |       |        | Mars | 13:07:20 |        |             |           |       |
| NIC2 | F215N | 2.1488 | Mars | 11:51:25 | 0.0200 | 2.58100e-21 | 0.0083705 | 0.864 |
|      |       |        | Sky  | 11:52:47 |        |             |           |       |
|      |       |        | Mars | 11:54:08 |        |             |           |       |
| NIC2 | F216N | 2.1642 | Mars | 12:00:54 | 0.0208 | 2.34700e-21 | 0.0083705 | 11.439 |
|      |       |        | Sky  | 12:02:16 |        |             |           |       |
|      |       |        | Mars | 13:03:36 |        |             |           |       |
| NIC2 | F237M | 2.3677 | Mars | 11:57:10 | 0.1546 | 2.42800e-22 | 0.0083705 | - |
|      |       |        | Sky  | 11:58:32 |        |             |           |       |
|      |       |        | Mars | 11:59:53 |        |             |           |       |

**Table II. Summary of total error introduced from error propagation**

| Error type | 1 SD Precision | 1 SD Accuracy |
|---|---|---|
| post-instrumental correction | 1% | 1% |
| conversion to I/F | 1% | 5% |
| Minnaert correction | 3% | 7% |
| Phobos/ISM bootstrap | 3% | 21% |
| Atmospheric Correction | 4% | 21% |
| Band depth retrieval | 4% | 21% |

**Table III.** Measured errors for band depth maps presented in Fig. 11.

| Band depth map wavelength (μm) | 1-standard deviation point-to-point variability in band depth |
|---|---|
| .970 | 0.0040 |
| 1.081 | 0.0020 |
| 1.130 | 0.0022 |
| 1.450 | 0.0024 |
| 1.900 | 0.0017 |

**Table IV.** Eigenvectors and their association to spectral features

|  | Explains % variance | Correlation with 0.950-μm reflectance map | | Correlation with 1.45-μm band depth map | | Correlation with 1.66-μm band depth map | | Correlation with 1.9-μm band depth map | |
|---|---|---|---|---|---|---|---|---|---|
|  |  | all | no cap | all | no cap | all | no cap | all | no cap |
| PC1 | 96.4 | 0.0902 | 0.921 | -0.322 | -0.520 | 0.0415 | -0.0080 | 0.362 | 0.390 |
| PC2 | 3.45 | 0.111 | -0.226 | 0.801 | 0.616 | 0.608 | 0.0510 | -0.422 | -0.408 |
| PC3 | 0.092 | 0.177 | 0.110 | 0.204 | 0.0392 | 0.433 | 0.231 | 0.460 | 0.617 |
| PC4 | 0.021 | -0.090 | -0.135 | 0.315 | 0.327 | -0.0438 | -0.255 | -0.421 | -0.449 |
| PC5 | 0.017 | -0.07 | -0.121 | -0.335 | -0.618 | 0.462 | 0.504 | 0.114 | 0.156 |
| PC6 | 0.0071 | 0.0410 | 0.039 | -0.0867 | -0.171 | 0.428 | 0.539 | -0.504 | -0.509 |

**Table V.** Interpretation of the eigenvectors (from correlation between eigenimages and surface geography)

| Principal Component | Geographical association | Interpretation |
|---|---|---|
| PC1 | All, except polar cap | Surface albedo variability |
| PC2 | Polar Cap, northern classical dark terrain, southern limb, southern hemisphere bright terrains | surface and atmospheric water ice, hydrous minerals |
| PC3 | Polar Cap, northern hemisphere | surface ice, atmospheric water vapor (?) |

**Figure Captions**

**Figure 1.**  Image taken through the HST/NICMOS F095N filter on 23 July 1997 ($L_s$ = 152°) showing Mars at a phase angle of 40°.  The central meridian longitude is near 35°W. The north polar cap is at top, Acidalia is the northern dark region, Arabia is the bright terrain on the Eastern limb, and the southern dark terrains include Margaritifer Terra and Sinus Meridiani.

**Figure 2.**  Mars data sets used for comparison and calibration purposes: a) original NICMOS data, in Mollweide equal-area map projection centered on 0° latitude and 0° longitude.  This projection is used for all subsequent mapped images.  b) Viking IRTM albedo map (Palluconi and Kieffer, 1981), c) HST/WFPC2 violet (0.255-µm) filter, d) MGS/MOLA altimetry (Smith *et al.,* 1999) , e) Odyssey/GRS epithermal neutron map (Feldman *et al.,* 2002).

**Figure 3.**  Representative spectrum taken by Phobos-2/ISM (*e.g.,* Murchie *et al.,* 1993) compared with NICMOS primary calibration (x), NICMOS bootstrapped calibration (*), and calibrated WFPC2 data of the same region on Mars. The Phobos-2 and NICMOS data were taken ~9 years apart, and the WFPC2 data were taken two weeks before the NICMOS data.

**Figure 4.**  1.45 µm reflectance scatter plot (NICMOS vs. ISM) for 6 different terrains.  Note that the NICMOS reflectivity is linearly correlated to the ISM reflectivity, justifying a correction for systematic errors by bootstrapping the NICMOS calibration to that of ISM.

**Figure 5.** Multiplicative factor (a, unitless) and additive factors (b, units of Minnaert-corrected I/F) required for each filter to match the NICMOS data to the Phobos/ISM data for the common surface regions studied.

**Figure 6.** Model transmission spectra for gaseous $CO_2$ (dashed) and $H_2O$ (dotted) for a single-layer, plane-parallel atmosphere at 275 K surface temperature and 8 mb surface pressure and a column density corresponding to a viewing angle of 45°. The $CO_2$ spectrum is offset downwards by 0.1. The positions of our NICMOS bandpasses are shown as asterisks.

**Figure 7.** Uncorrected NICMOS spectra of Southern dark terrain (x) compared to model atmospheric spectra (dotted line) described in the paper, convolved to NICMOS bandpasses (diamonds). Note the strong correlation between the model and observation for the absorption longward of 1.8 μm.

**Figure 8.** Curve used to determine the best scaling factor for the atmospheric pressure in our gaseous atmospheric radiative transfer model. The best fit is located at 0.45 times the calculated surface pressure.

**Figure 9.** Calculated single scattering albedo for water ice (solid line) and martian dust (dashed line). Asterisks show the spectra convolved to our NICMOS bandpasses. Note that whereas the water ice will generate a detectable band depth at 2.12-μm, martian dust as modeled here does not affect our band depth analysis in any detectable way.

**Figure 10.** Scatter plot of the 1.4 vs. 1.9 µm band depths of mineral spectra convolved to the NICMOS bandpasses. Anhydrous minerals (black) typically concentrate around near-zero 1.4- and 1.9-µm band-depth, whereas hydroxides (red) tend to extend along different 1.4-µm band-depths and low 1.9-µm band depth. Minerals containing $H_2O$ (magenta and green) exhibit an increase in both band depths.

**Figure 11.** Band depth maps generated from the 1997 NICMOS data. Band depths are calculated as 1-(observed I/F / interpolated I/F), and in this work represent the depth of the "band" relative to a linear continuum defined by the two adjacent filters. (a) 0.97 µm albedo map of Mars, generated from our observations; (b) 0.97 µm band depth relative to a continuum defined between 0.95 and 1.08 µm; (c) 1.08 µm band depth relative to a continuum defined between 0.97 and 1.13 µm; (d) 1.13 µm band depth relative to a continuum defined between 0.97 and 1.45µm; (e) 1.45 µm band depth relative to a continuum defined between 1.13 and 1.66 µm; (f) 1.66µm band depth relative to a continuum defined between 1.45 and 1.90 µm; (g) 1.90 µm band depth relative to a continuum defined between 1.66 and 2.15µm

**Figure 12.** Band depth mapping scatter plots and maps for 1.45- vs 1.9-µm. (a and d), and for 1.45- vs 1.66-µm (b and e). The band depth images are color-coded to the colored boxes in the respective plots. Note the detached section at a 1.9-µm band depth value of 0.035 (magenta), having a higher 1.45-µm band depth than the main cluster, and its association to the northern part of Acidalia (also magenta). Also note the region of lowest 1.45-µm band-depth and its association with the classical intermediate terrain found southeast and southwest of Acidalia.

**Figure 13.** Eigenimages of the first 6 eigenvectors of a Principal Components Analysis transform applied to the 1997 NICMOS data.  a) 0.970-μm albedo b-g) 1st through 6th eigenimages.  Note the geographic correlation between the features of eigenimages 1 through 5 with the 970-μm albedo map, and the absence of any defined features in the 6th eigenimage.  Particularly noticeable are the north polar cap, which is highlighted in the 2nd and 3rd eigenimages, and the strong contrast between Acidalia, classical intermediate albedo terrain adjacent to southeast and southwest Acidalia, and the rest of the terrains surrounding these in the 5th eigenimage.

**Figure 14.** The first two eigenvectors of our PCA study.  a) The first eigenvector lacks strong features and may be associated through the first eigenimage to classic albedo variations.  b) The second eigenvector is plotted along with the spectrum of 100-μm ice and an ice block (Roush *et al.,* 1990).  The spectral shape of the second eigenvector, along with its geographical association with the eigenimage, indicate primarily fine-grained (~100-μm) surface ices in the north polar cap as well as ice in the southern hemispheric polar hood/cap (see Fig. 14b).  The rest of the eigenvectors could not be directly associated with other specific surfaces phases in this way.

**Figure 15**. Results from n-dimensional PCA analysis:  a) 3-d scatter plot of the first 3 eigenvectors, projected onto a plane so as to show maximum endmember contrast.  The axes have been labeled with the corresponding eigenvectors.  Note the color-coded lobes that allow us to identify the different endmembers.  b) Mollweide projection of the 0.970-μm NICMOS observation.  The color coded regions from (a) are overplotted on the data.  c) Spectra of selected surface endmembers, corresponding to intermediate terrain (blue), North Polar Cap (red), southern dark terrain (pink), northern Acidalia (yellow), and southern polar hood (cyan).

**Figure 16.** Scatter plot relating the band depths of laboratory-measured minerals to the observed band depths of Mars (blue). Symbols: Anhydrous minerals (black); hydroxides (red); hydrates (magenta and green). The 1-$\sigma$ error cross corresponds to the precision to which we know the location of each NICMOS pixel with respect to each other. Note that even with this error, the cluster corresponding to northern Acidalia and enclosed in a magenta outline in Figures 8 and 9 remains detached, and its location in the scatter plot is consistent with a mixture of hydrous and anhydrous materials.

**Figure 17.** Comparison of lab mineral spectra to the spectra of the endmembers identified from PCA. Except for (c), the spectra of the endmembers have been multiplied by 10 to enhance the contrast, and all spectra have been selectively shifted for clarity. a) The spectrum of northern Acidalia compared to the spectra of both hydrous and anhydrous minerals having positive 1.45- and 1.90-$\mu$m band-depths, and a negative 1.66-$\mu$m band-depth. b) The spectrum of the southern dark terrain compared to the spectra of hydrous and anhydrous minerals having no absorption at 1.45-$\mu$m, a negative band-depth at 1.66-$\mu$m, and a positive band-depth at 1.90-$\mu$m. c) The spectrum of the north polar cap, compared to the spectrum of fine-grained, and coarse water ice. d) The spectrum of intermediate terrain compared to the spectra of both hydrous and anhydrous minerals having negative 1.45- and 1.66-$\mu$m band-depths, and positive 1.9-$\mu$m band-depths.

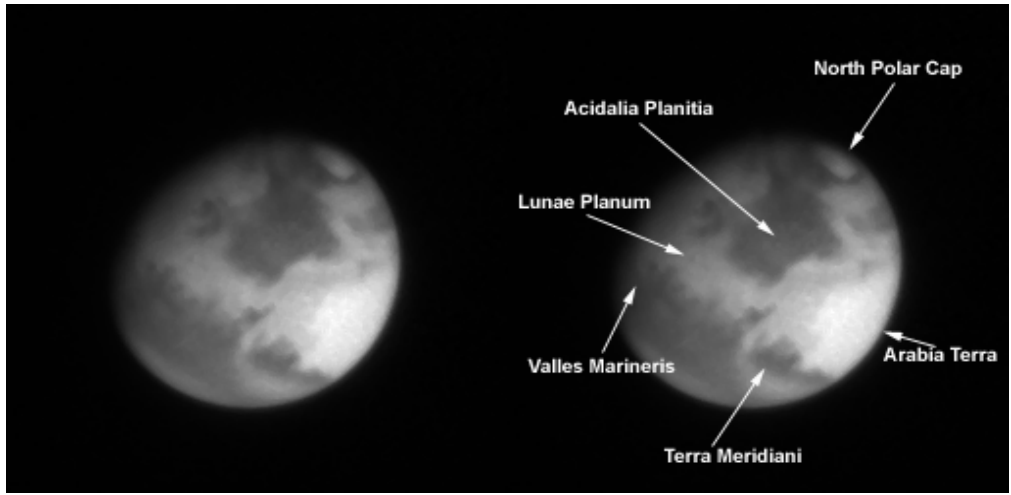

Figure 1

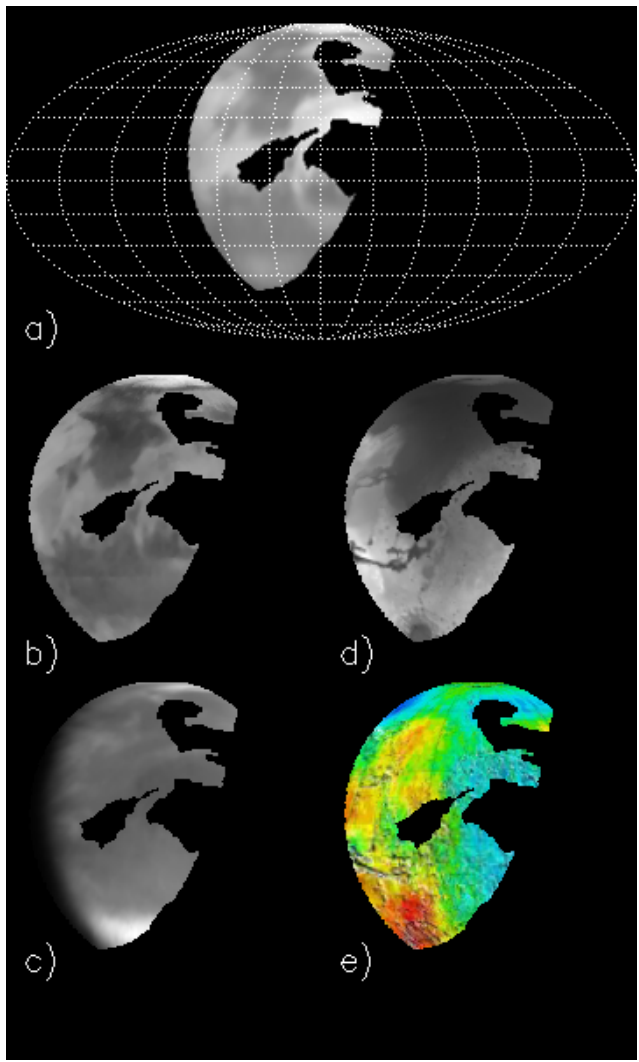

Figure 2

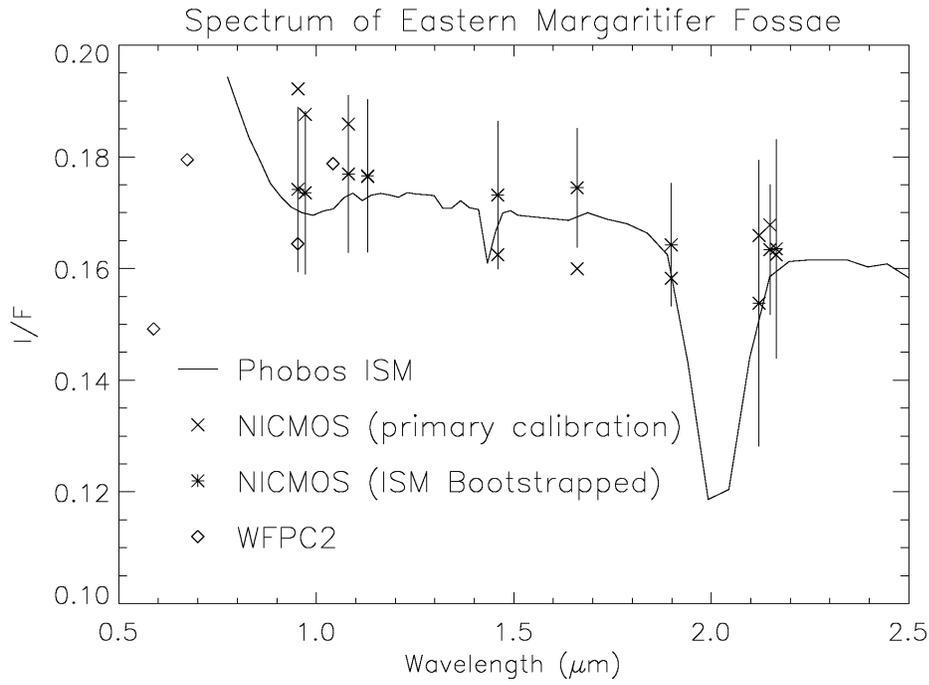

Figure 3

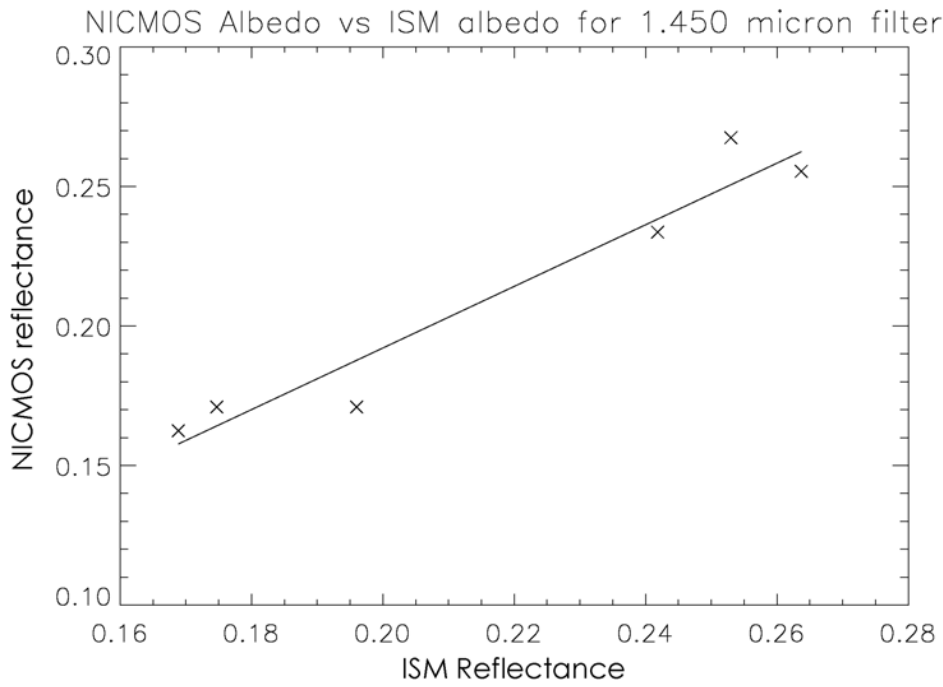

Figure 4

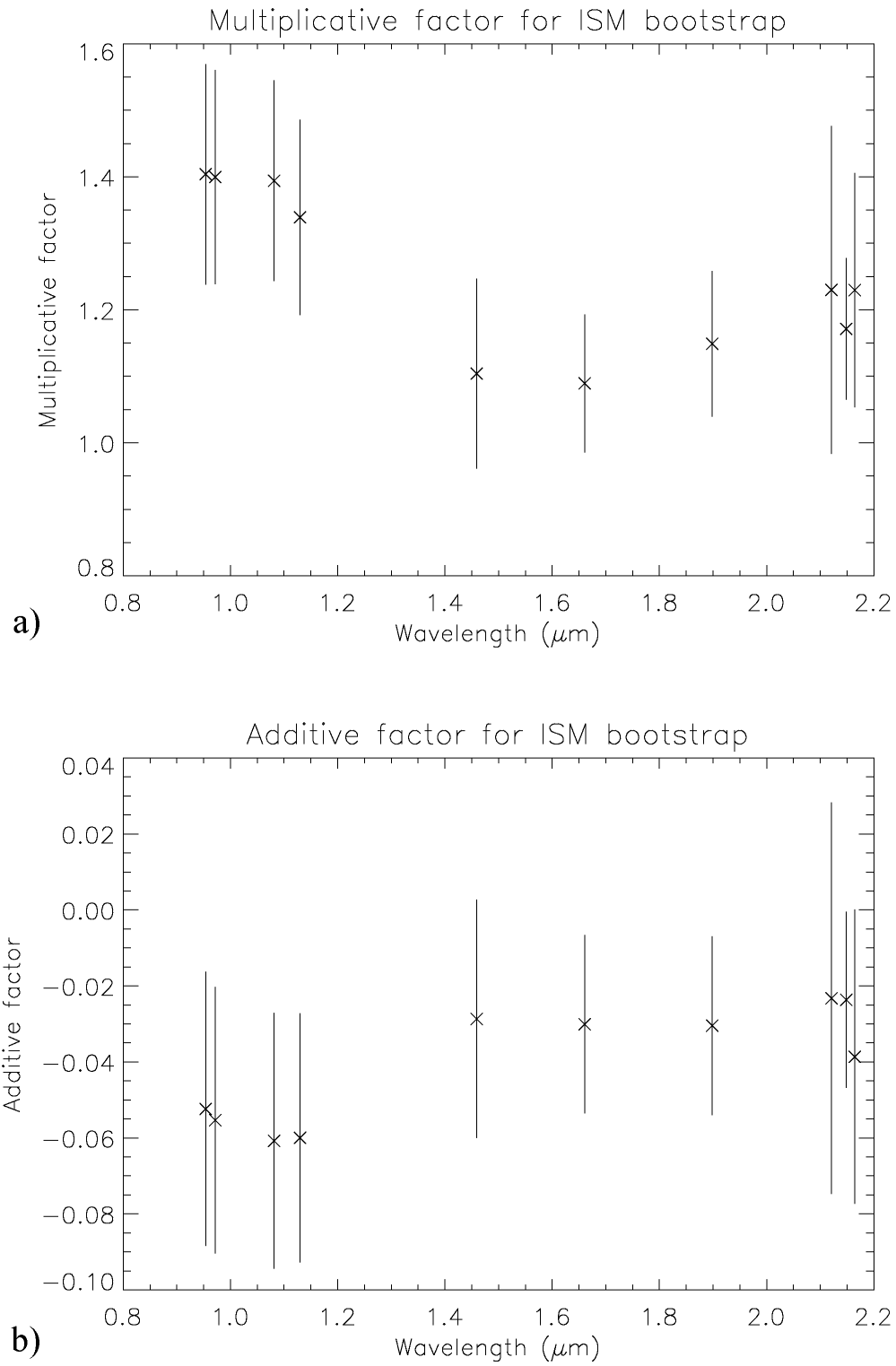

Figure 5

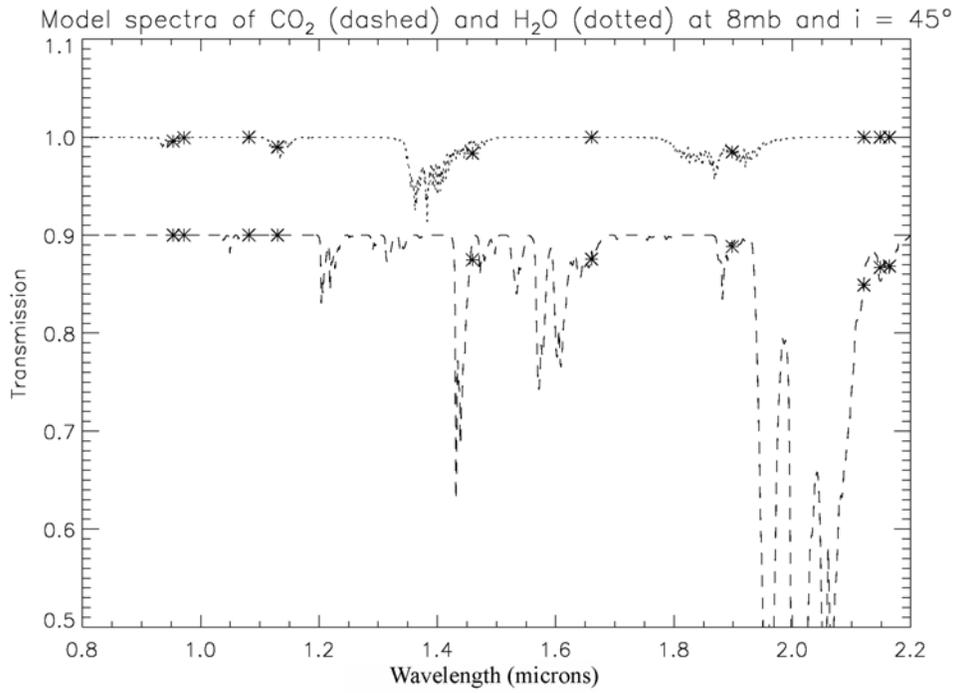

Figure 6

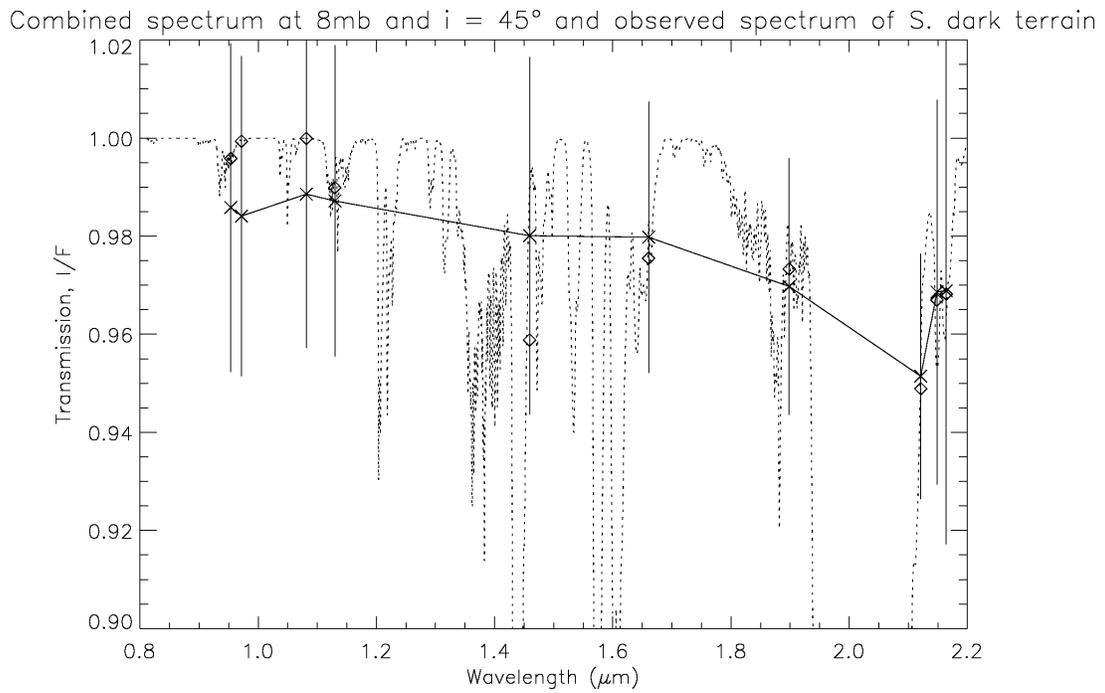

Figure 7

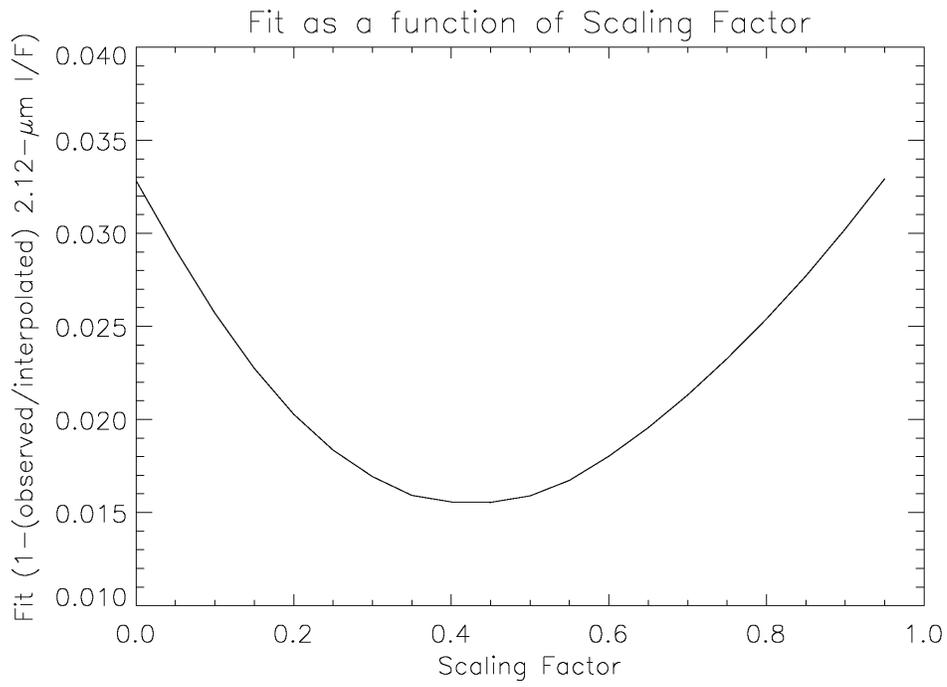

Figure 8

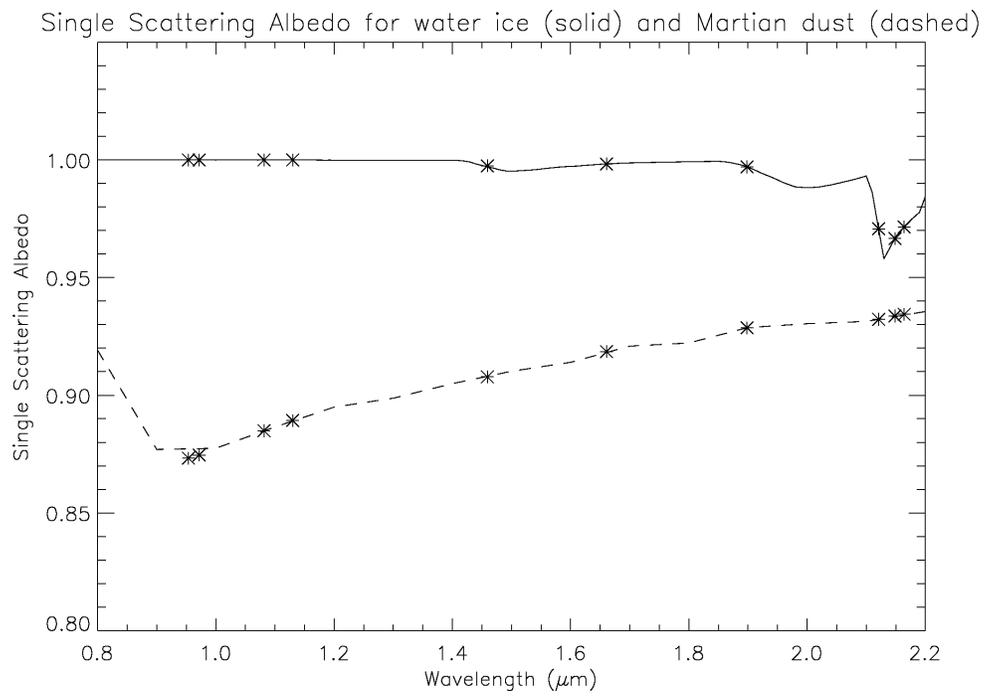

Figure 9

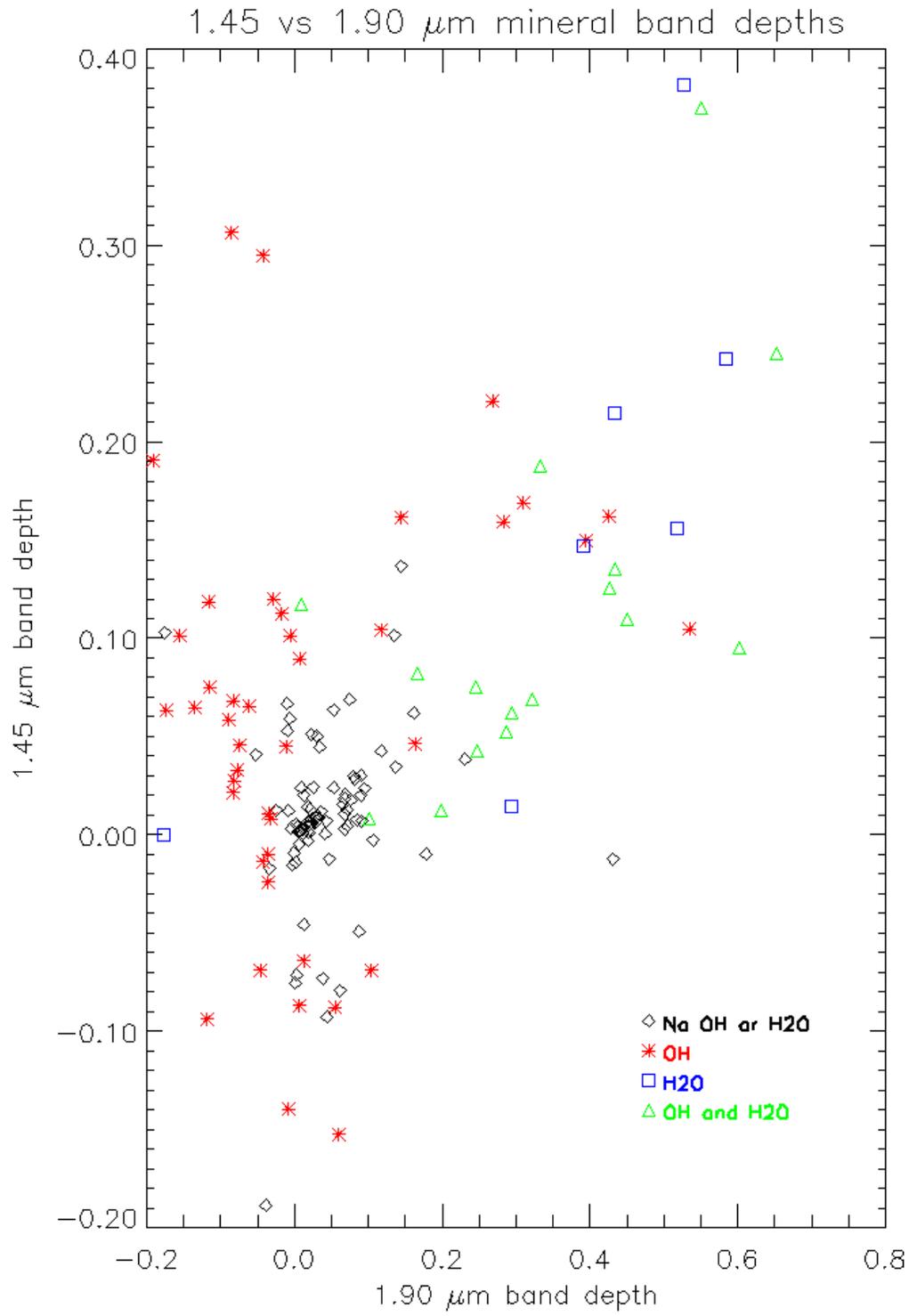

Figure 10

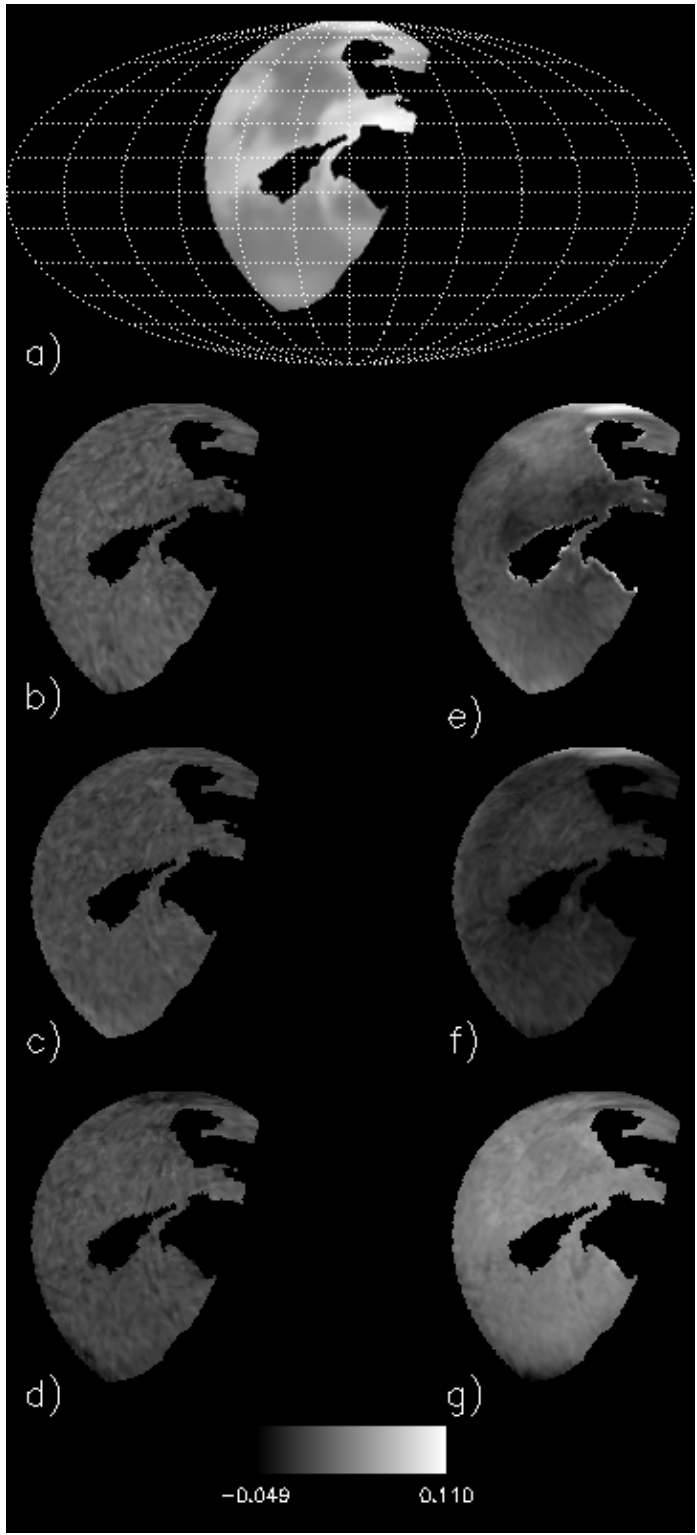

Figure 11

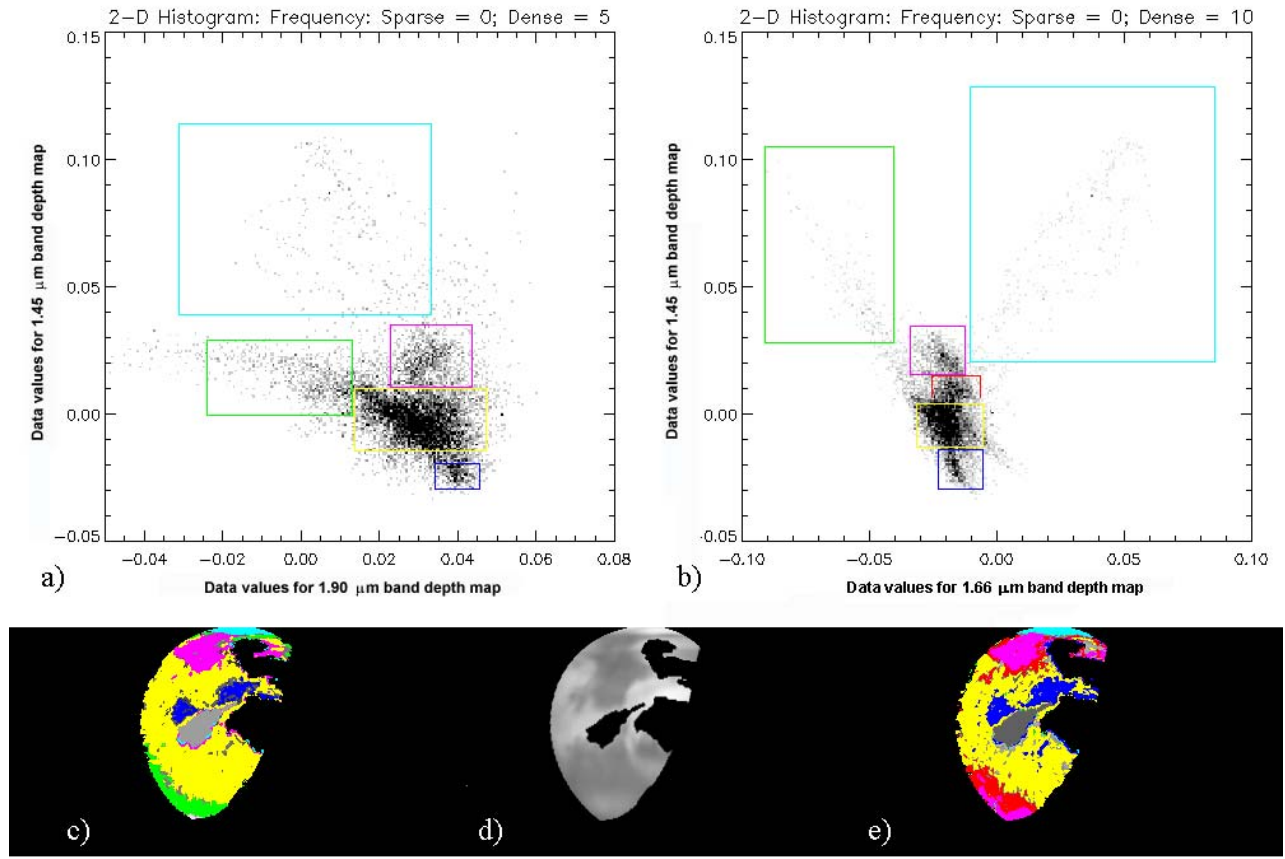

Figure 12

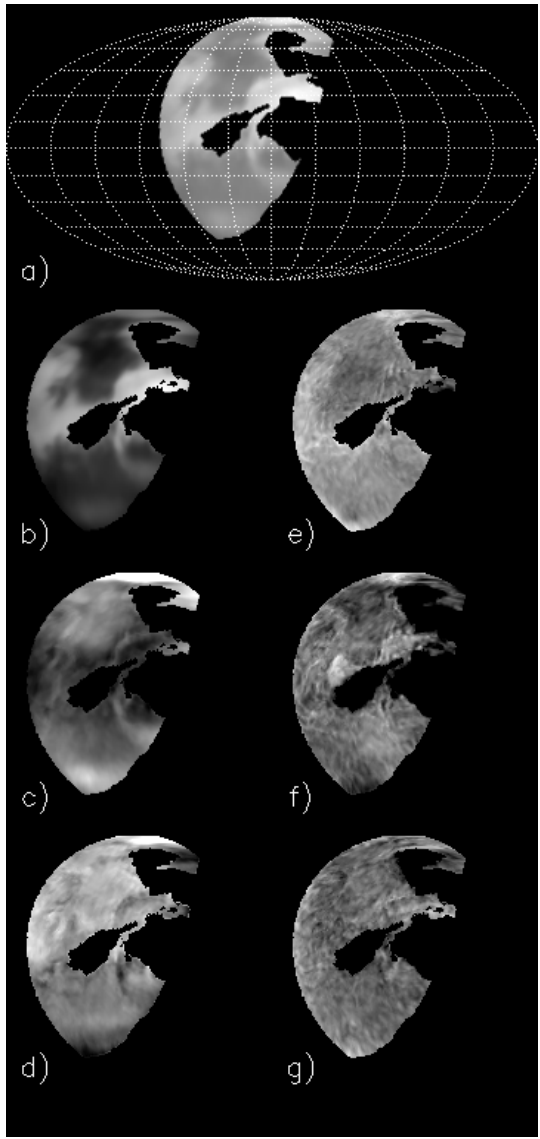

Figure 13

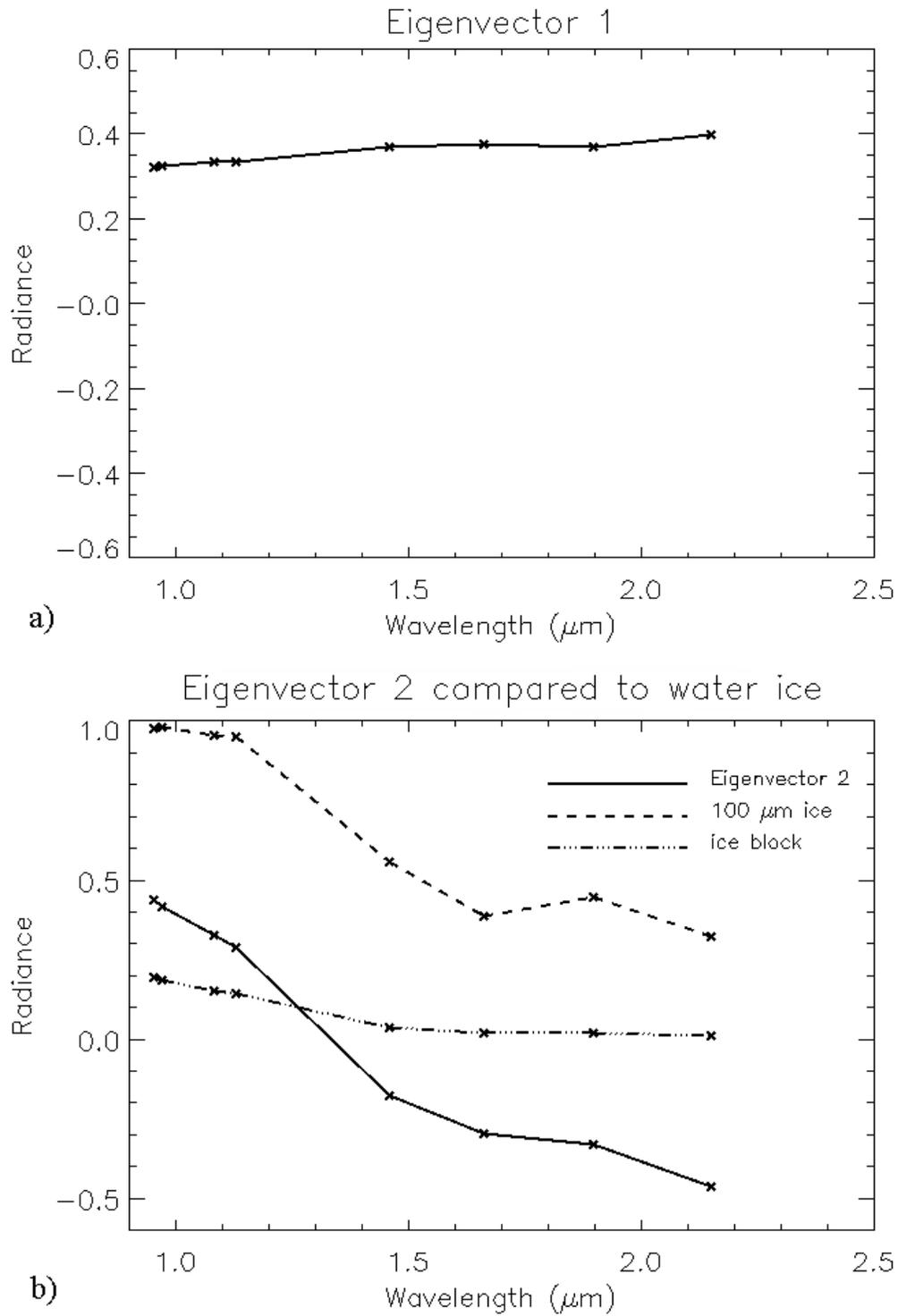

Figure 14

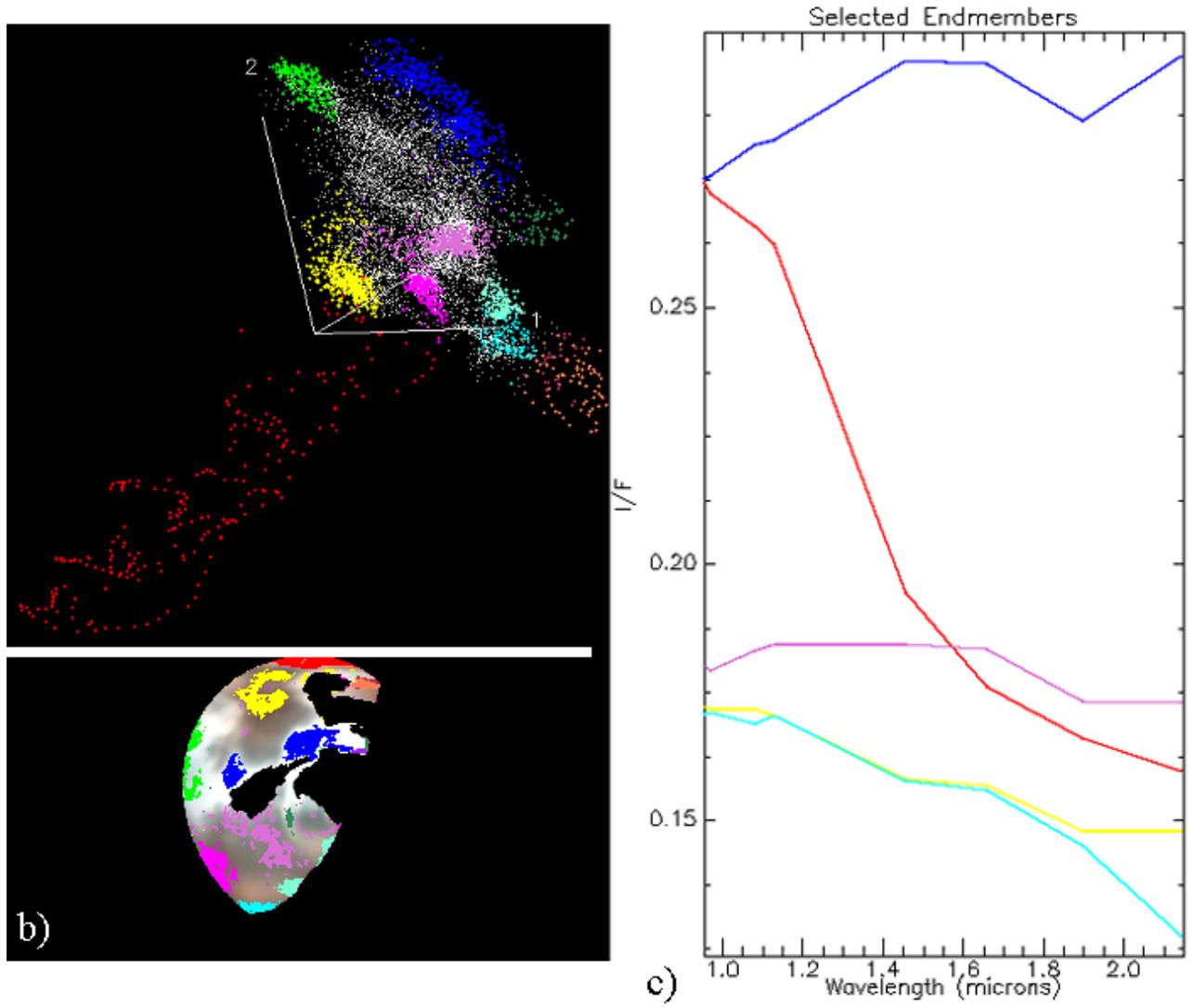

Figure 15

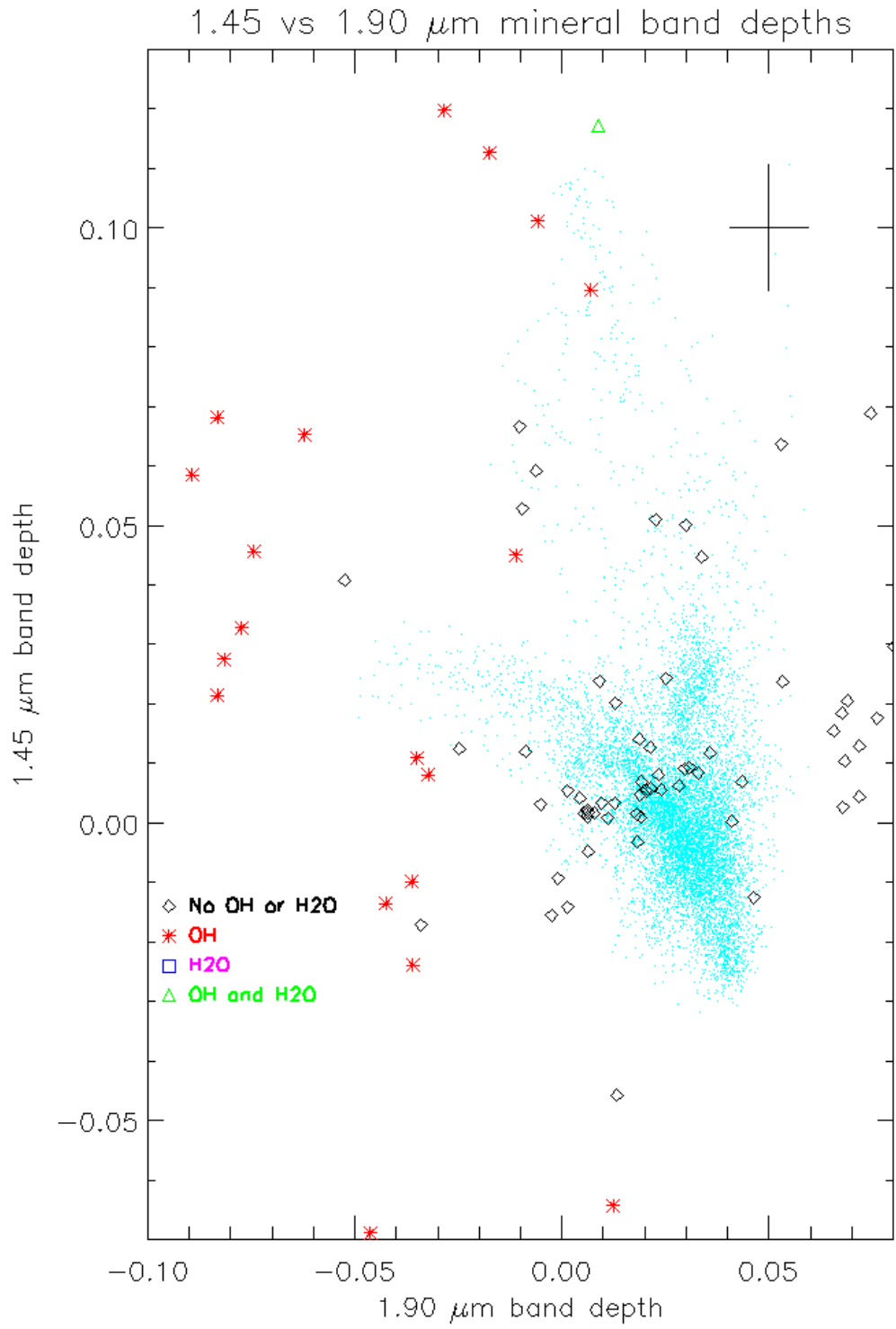

Figure 16

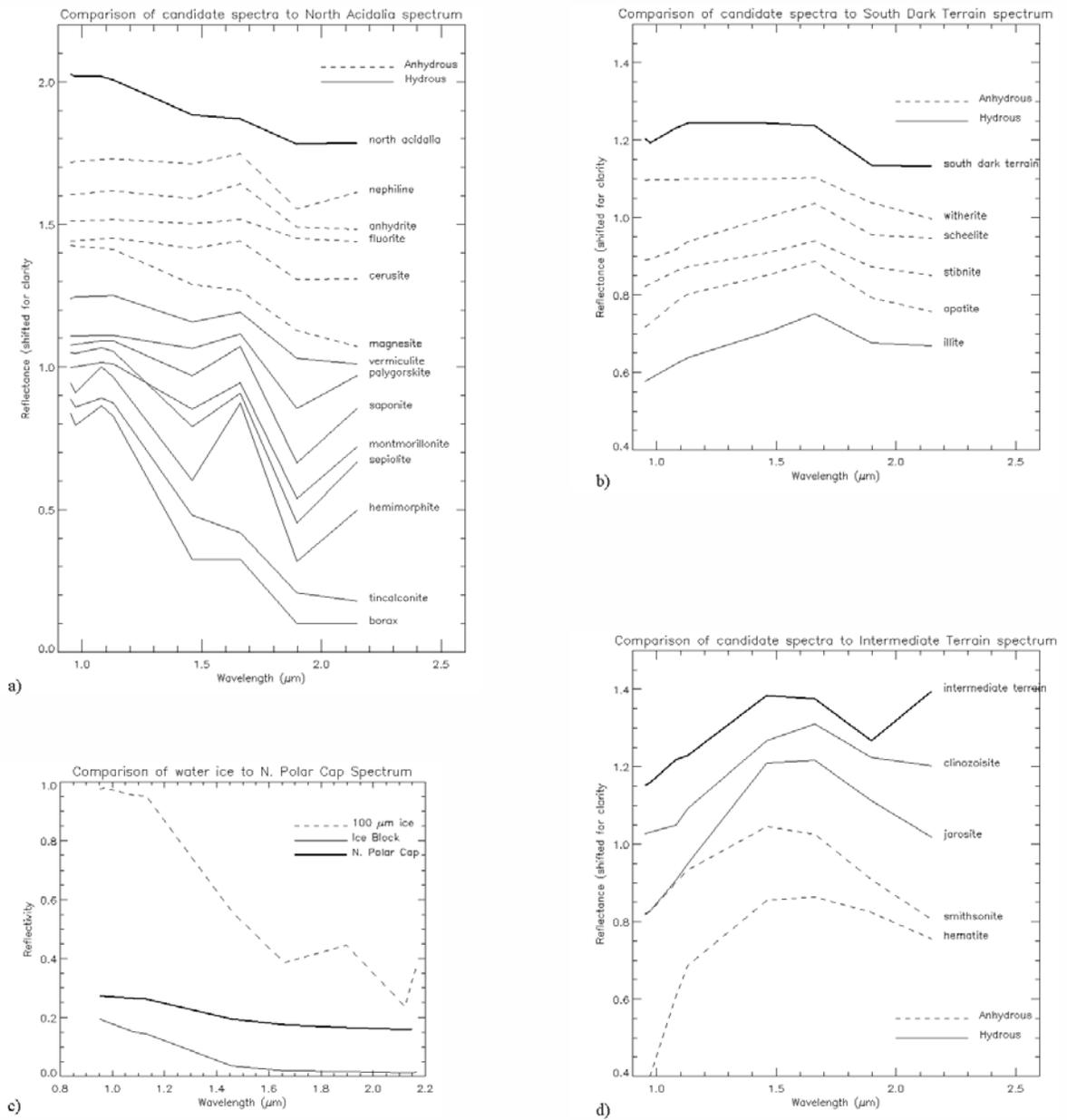

Figure 17